\documentclass[structabstract,printer]{aa}  

\usepackage{graphicx}
\usepackage{txfonts}
\usepackage{natbib}
\usepackage{lscape}
\bibpunct{(}{)}{;}{a}{}{,}

\begin{document}

\title{A single-dish survey of the HCO$^{+}$, HCN, and CN emission toward the T~Tauri disk population in Taurus}   
\subtitle{} 

\author{D.M. Salter 
\and 
M.R. Hogerheijde 
\and
R.F.J. van der Burg 
\and
L.E. Kristensen 
\and
C. Brinch 
}

\institute{Leiden Observatory, Leiden University, P.O. Box 9513, 2300 RA Leiden, The Netherlands \\
\email{demerese@strw.leidenuniv.nl} } 

\date{Submitted; accepted.}

\authorrunning{Salter et al.} 

\titlerunning{HCO$^{+}$, HCN, and CN emission toward the T~Tauri disk population in Taurus} 

\abstract
{The gas and dust content of protoplanetary disks evolves over the course of a few Myr. As the stellar X-ray and UV light penetration of the disk depends sensitively on the dust properties, trace molecular species like HCO$^+$, HCN, and CN are expected to show marked differences from photoprocessing effects as the dust content in the disk evolves.} 
{We investigate specifically the evolution of the UV irradiation of the molecular gas in protoplanetary disks around a sample of classical T~Tauri stars in Taurus that exhibit a wide range in grain growth and dust settling properties.} 
{We obtained HCO$^+$\,($J$\,=\,3--2), HCN\,($J$\,=\,3--2), and CN\,($J$\,=\,2--1) observations of 13 sources with the James Clerk Maxwell Telescope. Our sample has 1.3\,mm fluxes in excess of 75\,mJy, indicating the presence of significant dust reservoirs; a range of dust settling as traced through their spectral slopes between 6, 13, and 25\,$\mu$m; varying degrees of grain growth as extrapolated from the strength of their 10-$\mu$m silicate emission feature; and complements data from the literature, essentially completing the molecular line coverage for the 21 brightest millimeter targets in the Taurus star-forming region. We compare the emission line strengths with the sources' continuum flux and infrared features, and use detailed modeling based on two different model prescriptions to compare typical disk abundances for HCO$^+$, HCN, and CN with the gas-line observations for our sample.} 
{We detected HCO$^+$\,(3--2) toward 6 disks, HCN\,(3--2) from 0 disks, and CN\,(2--1) toward 4 disks, with typical 3$\sigma$ upper limits of 150--300\,mK ($T_{\rm mb}$) in 0.2\,km\,s$^{-1}$ channels. For the complete sample, there is no correlation between the gas-line strengths or their ratios and either the sources' dust continuum flux or infrared slope. Modeling shows that fractional abundances of 10$^{-9}$--10$^{-11}$ can explain the line intensities observed, although significant modeling uncertainties remain. For DG Tau, which we model in great detail, we find that two very different temperature and density disk structures produce very similar lines for the same underlying abundances. Instead, it is the gas properties (particularly $T_{\rm kin}$ and $R_{\rm gas}$) and the projected kinematics (determined by M$_\star$\,sin\,$i$) that have the largest impact on line appearance; and these system parameters are not well-constrained by current dust models, but instead will be probed directly with ALMA. } 
{Unresolved observations of the dust continuum provide neither a unique nor a complete picture of protoplanetary disks. Instead, gas-line measurements and resolved observations of dust and gas alike are needed to arrive at a full picture.}

%


\maketitle


\section{Introduction}

Many young pre-main-sequence (PMS) stars of a few solar masses or less (T~Tauri stars) are surrounded by disks of gas and dust in the midst of forming planetary systems \citep{beckwith1996,dutrey2007}. Observations with new instruments in the sub-millimeter and infrared wavelength regimes have provided a wealth of information about the dust content in these disks. Physical models are now fitted successfully to the observed spectral energy distributions (SEDs), revealing details on their large-scale radial and vertical structure \citep[e.g.][]{calvet2002}. Meanwhile, the millimeter spectral slope and the silicate emission features at 10 and 20\,$\mu$m indicate the amount of small-scale dust processing, including grain growth \citep[e.g.][and references therein]{natta2007}. These dust properties determine the overall opacity of the disk. A broad range in properties, in turn, implies different self-shielding abilities from the strong X-ray and ultra-violet (UV) radiation fields emitted by the central star. As this high-energy radiation penetrates the disk more or less efficiently, possibly irradiating deeper disk layers or a greater proportion of the available gas reservoir, then any molecular gas content that is left unprotected by the dust is subject to increased photoionization and photodissociation processes.   

One key development in the evolution of disks is the growth of dust particles from sub-micron sizes first to larger sizes, and ultimately planets. The SEDs of many disks exhibit clear evidence of this growth to micron and even centimeter sizes \citep[e.g.][]{beckwith1991,testi2003}. When destructive collisions are negligible, this grain growth process inherently depletes the small-grain population, which is the dominant source of UV opacity in the disk. In addition, as the dust particles grow, they are expected to decouple from the gas and settle to the disk midplane \citep{weidenschilling1997,dullemond2004}, leaving the gas more exposed to both X-ray and UV irradiation in the upper disk layers. In a recent analysis of Spitzer IRS data for a large sample of T~Tauri stars, \citet{furlan2006} find evidence of this sedimentation of dust particles to the midplane. They conclude from the infrared spectra that the small dust content has already decreased by 2--3 orders of magnitude in the surface layers of many disks in the Taurus star-forming region. Since the processes of grain growth and dust removal (via settling) both correspond to a lower disk opacity, we can now test the link between the dust properties and the photoprocessing effects on the molecular gas composition.

Observations of molecular lines are required to probe the chemical content and evolution of the molecular gas. To date, only about three dozen disks have been detected successfully in rotational lines of CO \citep{koerner1995,mannings1997,mannings2000,dutrey1997,guilloteau1999,simon2000,thi2001,kempen2007,pietu2007,chapillon2008,schreyer2008,schaefer2009}, and only a handful of disks have been searched for rarer species like HCO$^+$, HCN, and CN \citep[e.g][]{dutrey1997,kesslerphd2004,thi2004,kastner1997,kastner2008a,kastner2008b}. The latter three molecules are indicative of the rates at which X-ray ionization and UV photodissociation is occurring in the intermediate layers of the disk. 

In those observational studies, HCO$^+$ and CN often represent two of the strongest emission lines detected toward disks around PMS stars. Models predict that the abundances of HCO$^+$, HCN, and CN should increase similarly as the X-ray ionization rate is increased \citep{lepp1996,glassgold2004}. However, the UV field then further enhances CN, a photodissociation product, at the expense of HCN, its predecessor \citep{lepp1996,aikawa2002}. In this way, we can use measurements of HCO$^+$ as a normalization factor for the total ionization (or overall irradiation) effect on the disk, while the ratio of CN to HCN probes the UV photodissociation process, which is most sensitive to grain growth and dust settling.

Young forming stars are the main contributors to the UV interstellar radiation field \citep{herbig1986}. In star-forming regions permeated by moderate UV radiation fields from newly forming stars, the ratio of CN to HCN is observed to be very strongly enhanced with respect to typical molecular cloud values \citep{fuente1993,hogerheijde1995a,greaves1996,terzieva1998}. The same shift is predicted by chemical disk models where the ratio of CN over HCN changes radically in favor of CN as the stellar UV flux increases \citep{zadelhoff2003}. Young stars also possess strong X-ray activity levels \citep{gudel2008a}. Therefore, the brightest HCO$^+$ lines and the largest CN to HCN ratios are expected from luminous sources and optically thin disks. 

Previously, \citet{kastner2008b} summarized the HCO$^+$, HCN, and CN line observations available in the literature for disks around PMS stars, showing how the chemical impact of the X-ray and UV fields is significant in the general disk population and that the line ratios do exhibit tentative trends that correlate with the expected photoprocessing effects. However, also evident is the small number of \emph{widely-differing} systems that have been studied in any detail at all. This makes it difficult to distinguish evolutionary differences from source-to-source variations in mass, luminosity, environment, and orientation. Naturally, larger homogeneous surveys are neccessary to identify gas-rich disks and to provide observational tests of the processes driving their physical and chemical evolution. 

Here we present a study of HCO$^+$, HCN, and CN line emission toward \emph{all} the classical T~Tauri stars in Taurus with a compact 1.3\,mm continuum flux $\ge$\,75\,mJy, or equivalently a disk mass $\gtrsim$\,0.014\,M$_\odot$, using the standard flux relation and adopting the same mean disk parameters as \citet{dutrey1996}. These parameters include a disk temperature of 15\,K, a dust absorption coefficient $K_{\rm 1.3mm}$ of 0.02\,cm$^2$\,g$^{-1}$, and a distance to Taurus of 140\,pc. Consequently, our selection criterion favors warmer and more massive disks in the star-forming region. 

The line data were obtained from new single-dish observations, as well as taken from the literature, where available. In Sect.~\ref{obs} we describe the details of the observations and in Sect.~\ref{results} we report the line intensities and detection limits obtained from our measurements and from the literature. Sect.~\ref{trends} investigates observational trends between the dust emission properties of the disk sample, such as millimeter flux and infrared slope, and the gas-line strengths or upper limits. Because none of these tracers spatially resolve the disks, and because line strengths depend both on the total amount of molecules and on the excitation conditions, Sect.~\ref{modeling} employs a modeling approach to constrain the HCO$^+$, HCN, and CN abundances. This is done using two generic models (Sect.~\ref{general}) and, for one source, DG Tau, in particular, a more detailed approach was taken (Sect.~\ref{dgtau}). Sect.~\ref{discussion} discusses our results and the limitations of our methods and Sect.~\ref{summary} summarizes the main findings.


\begin{table*}[tbh!] 
\begin{minipage}[t]{1.0\linewidth} 
\caption{Taurus disk sample. \label{sources} }  
\centering 
\begin{tabular}{lcccccccccc} 
\hline 
\hline 
\noalign{\smallskip}

(1) & (2) & (3) & (4) & (5) & (6) & (7) & (8) & (9) & (10) & (11) \\
Source & RA & DEC & F$_{\rm 1.3\,mm}$ & $\Delta n$ & 10$\rm \mu$m & M$_\star$ & Sp. & H$\alpha$\,EW & $L_{\rm bol}$ & $L_{\rm X}$ \\ 
Name & (J2000) & (J2000) & [mJy] & ~ & Strength & [M$_\odot$] & Type & [$\AA$] & [L$_\odot$] & [10$^{30}$\,erg\,s$^{-1}$] \\

\hline
\multicolumn{11}{c}{\textit{Our subsample with new molecular line observations presented here}}\\
\hline

\object{CW\,Tau}   &  04:14:17.00  &  $+$28:10:57.8  &  96\,$\pm$\,8    &  0.96  &  0.16  &  1.06  &  K3    &  135  &  1.35  &  2.844       \\
\object{CY\,Tau}   &  04:17:33.72  &  $+$28:20:46.8  &  133\,$\pm$\,11  &  0.50  &  0.12  &  0.55  &  M1.5  &  70   &  0.47  &  0.163       \\
\object{FT\,Tau}   &  04:23:39.18  &  $+$24:56:14.3  &  130\,$\pm$\,14  &  0.31  &  0.43  &  -     &  M3    &  -    &  $>$\,0.38  &  0.799       \\
\object{DG\,Tau}   &  04:27:04.70  &  $+$26:06:16.2  &  443\,$\pm$\,20  &  0.44  &  0.00  &  0.67  &  K6    &  113  &  0.90  &  $<$\,0.766  \\
\object{IQ\,Tau}   &  04:29:51.56  &  $+$26:06:44.8  &  87\,$\pm$\,11   &  0.40  &  0.35  &  0.68  &  M0.5  &  8    &  0.65  &  0.416       \\
\object{V806\,Tau} &  04:32:15.41  &  $+$24:28:60.0  &  124\,$\pm$\,13  &  1.46  &  0.35  &  -     &  M0    &  88   &  1.30  &  -	       \\
\object{UZ\,Tau}   &  04:32:42.96  &  $+$25:52.31.1  &  172\,$\pm$\,15  &  0.58  &  0.45  &  0.18  &  M1-2  &  82   &  1.06  &  0.890       \\
\object{CI\,Tau}   &  04:33:52.01  &  $+$22:50:30.1  &  190\,$\pm$\,17  &  0.93  &  0.54  &  0.77  &  K7    &  102  &  0.87  &  0.195       \\
\object{DN\,Tau}   &  04:35:27.37  &  $+$24:14:59.6  &  84\,$\pm$\,13   &  0.33  &  0.22  &  0.70  &  M0    &  12   &  0.91  &  1.155       \\
\object{DO\,Tau}   &  04:38:28.58  &  $+$26:10:49.9  &  136\,$\pm$\,11  &  0.64  &  0.18  &  0.70  &  M0    &  100  &  1.20  &  $<$\,0.270  \\
\object{GO\,Tau}   &  04:43:03.08  &  $+$25:20:18.6  &  83\,$\pm$\,12   &  0.97  &  0.40  &  0.71  &  M0    &  81   &  0.27  &  0.249       \\
\object{DQ\,Tau}   &  04:46:53.06  &  $+$17:00:00.2  &  91\,$\pm$\,9    &  0.19  &  0.14  &  0.43  &  M0    &  113  &  0.91  &  -	       \\
\object{DR\,Tau}   &  04:47:06.22  &  $+$16:58:42.9  &  159\,$\pm$\,11  &  0.61  &  0.27  &  0.38  &  K5    &  87   &  3.00  &  -	       \\

\hline
\multicolumn{11}{c}{\textit{Sources with molecular line data already in the literature (see Table \ref{fluxdata}) }} \\
\hline

\object{V892\,Tau} &  04:18:40.62  &  $+$28:19:15.5  &  289\,$\pm$\,13  &  -1.22  &  0.65   &  -     &  B9    &  13   &  20.90  &  8.576      \\
\object{RY\,Tau}   &  04:21:57.41  &  $+$28:26:35.6  &  229\,$\pm$\,17  &  0.46	&  1.36   &  1.63  &  K0    &  20   &  7.60   &  5.520      \\
\object{T\,Tau}    &  04:21:59.43  &  $+$19:32:06.4  &  280\,$\pm$\,9   &  1.19	&  -0.10  &  -     &  K7    &  60   &  15.50  &  8.048      \\
\object{GG\,Tau}   &  04:32:30.41  &  $+$17:31:40.2  &  593\,$\pm$\,53  &  0.74	&  0.64   &  1.55  &  K7    &  54   &  1.55   &  0.160      \\
\object{DL\,Tau}   &  04:33:39.06  &  $+$25:20:38.2  &  230\,$\pm$\,14  &  0.24	&  0.09   &  0.70  &  K7    &  105  &  0.70   &  $<$\,1.000 \\
\object{DM\,Tau}   &  04:33:48.73  &  $+$18:10:10.0  &  109\,$\pm$\,13  &  1.93	&  0.96   &  0.25  &  M1.5  &  139  &  0.25   &  $<$\,0.457 \\
\object{AA\,Tau}   &  04:34:55.42  &  $+$24:28:53.2  &  88\,$\pm$\,9    &  0.50	&  0.35   &  0.78  &  K7    &  37   &  0.74   &  0.690      \\
\object{GM\,Aur}   &  04:38:21.34  &  $+$26:09:13.7  &  253\,$\pm$\,12  &  3.19	&  1.19   &  1.00  &  K7    &  96   &  1.00   &  1.241      \\

\hline
\noalign{\smallskip}
\end{tabular}
\end{minipage}

\hfill

\begin{minipage}[h]{\linewidth} 
\begin{tiny}
\textbf{Notes.} (Cols.~1-3) Source name and 2MASS coordinates; (Col.~4) 1.3\,mm continuum fluxes from \citet{beckwith1996}; (Cols.~5-6) Dust settling parameters $\Delta n$ and the 10$\mu$m silicate strength from \citet{furlan2006}; (Col.~7) Stellar masses from \citet{white2001} or \citet{hartigan1995}; (Col.~8) Spectral types from \citet{luhman2010} or \citet{rebull2010}; (Col.~9) H$\alpha$ equivalent widths from \citet{herbig1988}; (Col.~10) Bolometric luminosities from \citet{palla2002}, except for DQ Tau and DR Tau, which come from \citet{kenyon1995}; (Col.~11) X-ray luminosities from \citet{gudel2007a} or \citet{damiani1995a}.
\end{tiny}
\end{minipage}

\hfill
\end{table*}

\section{Observations}
\label{obs}

\subsection{The Taurus disk sample}
\label{sample}

The sample consists of the classical T~Tauri stars in the Taurus star-forming region that were included in the Spitzer IRS classification work by \citet{furlan2006} and possess a 1.3\,mm continuum flux larger than 75\,mJy \citep{beckwith1990}. The sources cover a broad range in spectral indices between 6, 13, and 25\,$\mu$m, and the shape and strength of their 10-$\mu$m silicate emission features. The difference in the mid-infrared spectral indices (defined as $\Delta n = n_{13-25} - n_{6-13}$) serves as an indicator of dust settling in the inner disk ($\leq$1\,AU, with smaller $\Delta n$ suggesting more settling), and the strength of the 10$\mu$m silicate feature decreases with a depleting population of small grains \citep{furlan2006}. This characterization of the current dust properties in the upper surface layers of the disk allows us to explicitly contrast the presence of small grains -- the main source of UV opacity -- with the line strengths and ratios of the molecular species that trace UV photodissociation. 

For better odds of a high signal-to-noise gas-line detection, we imposed the 75\,mJy continuum flux minimum cutoff. This criterion results from the millimeter flux being related to the dust mass, and the idea that a larger dust mass is indicative of a potentially large gas reservoir. Our source list includes some of the brightest young stellar objects (YSOs) in Taurus at millimeter wavelengths, many of which already possess HCO$^+$, HCN, and CN rotational line detections. Consequently, the cutoff value was chosen to be representative of the lowest circumstellar dust mass found in the literature for a disk that had been detected in at least one of the three molecules. Therefore, our study is not necessarily probing a lower dust-mass regime, but rather contributes to completing the sample of sources down to this flux cutoff, and expanding the subgroup of disks with similar dust masses that can be compared.

We restricted the sample to one star-forming region, so that our results are relatively insensitive to the initial cloud conditions. In addition, due to their shared environment, we can assume that the local interstellar radiation field is similar for all disks, and more importantly, that it plays a less significant role in irradiating the disk in the UV regime. It is generally assumed that the external contribution to UV photodissociation of the upper and outer disk layers is roughly 3--4 magnitudes lower in intensity than the stellar excess UV spectrum for T~Tauri stars, based on observations of five protostars by \citet{herbig1986}, of which four were located in Taurus. 

The full sample consists of 21 protoplanetary disks, including 13 that had not been observed previously in any of our molecular tracers. Complementery molecular line observations for those 13 sources are presented here. Table \ref{sources} lists their 2MASS source coordinates (used for our JCMT observations), 1.3\,mm continuum fluxes \citep{beckwith1990}, mid-infrared dust settling parameters \citep{furlan2006}, as well as several stellar properties that are closely tied to the radiation field.


\subsection{JCMT single-dish millimeter observations}
\label{jcmtobs}

Using the RxA receiver and the ACSIS backend on the James Clerk Maxwell Telescope (JCMT) on Mauna Kea, we observed the HCO$^{+}$\,$J$\,=\,3--2 (267.558\,GHz), HCN\,$J$\,=\,3--2 (265.886\,GHz), and CN\,$J$\,=\,2--1 hyperfine line (centered at 226.800\,GHz) emission toward the subsample of 13 sources. The observations were made between September 2007 and December 2009.\footnote{JCMT observing programs M07BN06 and M08BN01 (PI: Salter).} The HCO$^{+}$ observations were carried out in position-switched mode with a reference position at a 900$''$ offset in azimuth. Although a $900''$ position switch is unlikely to be off the Taurus molecular clouds, the observed lines trace relatively dense gas and little emission is expected away from the source. For the less abundant HCN and CN species, we used the more time-efficient beam-switching mode with a reference position at a throw of 180$''$. 

All observations had a bandwidth of 250\,MHz, giving a channel resolution of 30.5\,kHz (or $\sim$0.034\,km\,s$^{-1}$ for HCO$^{+}$ and HCN, and $\sim$0.040\,km\,s$^{-1}$ for CN). The atmospheric opacity was between 0.05 and 0.12 during the observations, with typical on-source integration times of 20 minutes (aside from one extra-deep 60-min observation of HCN toward the brightest source, DG Tau, in order to establish a deeper upper limit for the subsample). 

The JCMT beam is 19$''$ at 265\,GHz and 22$''$ at 226\,GHz, much larger than any protoplanetary disk at a distance of 140\,pc to the Taurus star-forming region \citep{kenyon1994}. Since many young stars are still situated near their parental cloud, any line detection needs to be checked for possible contamination from the extended cloud. Following the strategy of \citet{thi2001} and \citet{kempen2007}, every strong detection of HCO$^{+}$ was followed up with observations 30$''$ east and 30$''$ north of the object.

\subsection{Data reduction}

\begin{figure*} 
\centering 
\includegraphics[width=10.5cm,angle=-90]{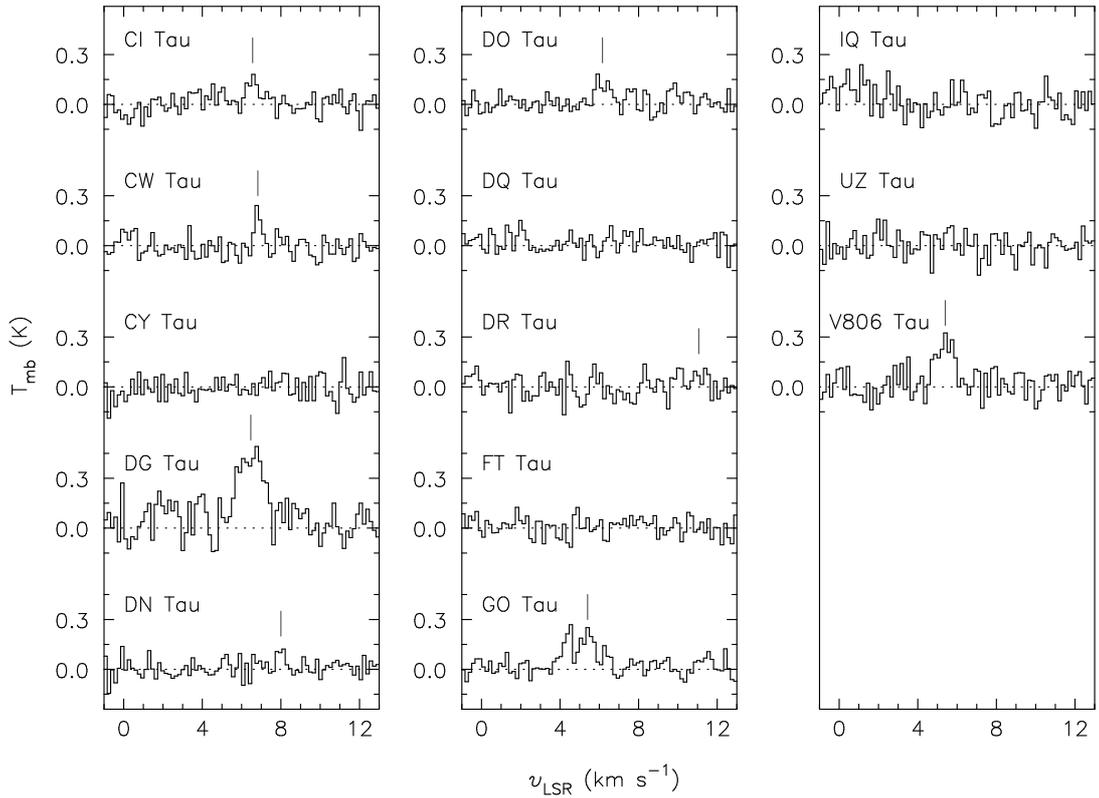} 
\caption{HCO$^{+}$\,($J$\,=\,3--2) observations of our 13-disk subsample in Taurus. A vertical line above the spectrum indicates the central radial velocity $\upsilon_{\rm LSR}$ for the detected emission, as determined from a Gaussian fit to the emission profile. In the case of the non-detection toward DR\,Tau, we indicate the expected $\upsilon_{\rm LSR}$ based on our fit to its CN emission profile in Fig.~\ref{cnfig}. \label{hcofig} }
\end{figure*}

\begin{figure*} 
\centering 
\includegraphics[width=10.5cm,angle=-90]{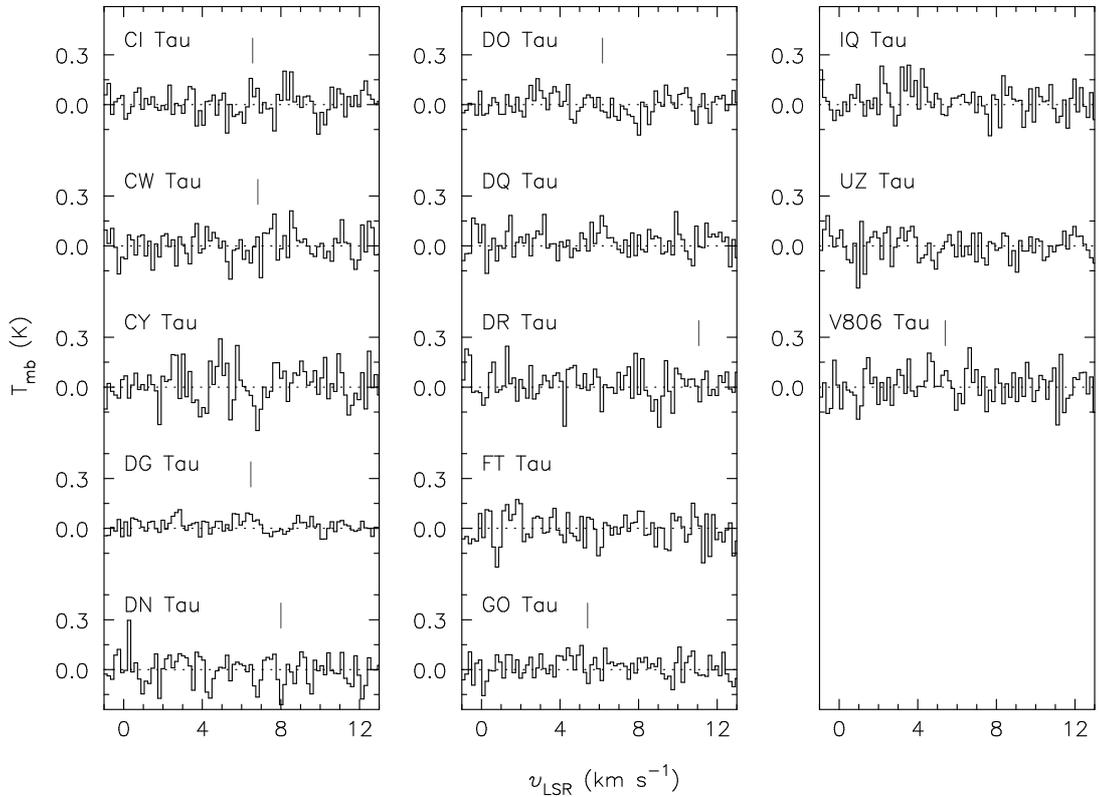}
\caption{HCN\,($J$\,=\,3--2) observations of our subsample in Taurus. A vertical line above the spectrum indicates the central velocity found from the detected HCO$^{+}$ emission in Fig.~\ref{hcofig}, except for DR\,Tau where the detected CN emission velocity is used instead. We note that HCN emission is not detected toward any of the sources, including the most massive disk DG\,Tau. \label{hcnfig} } 
\end{figure*}

\begin{figure*} 
\centering 
\includegraphics[width=10.5cm,angle=-90]{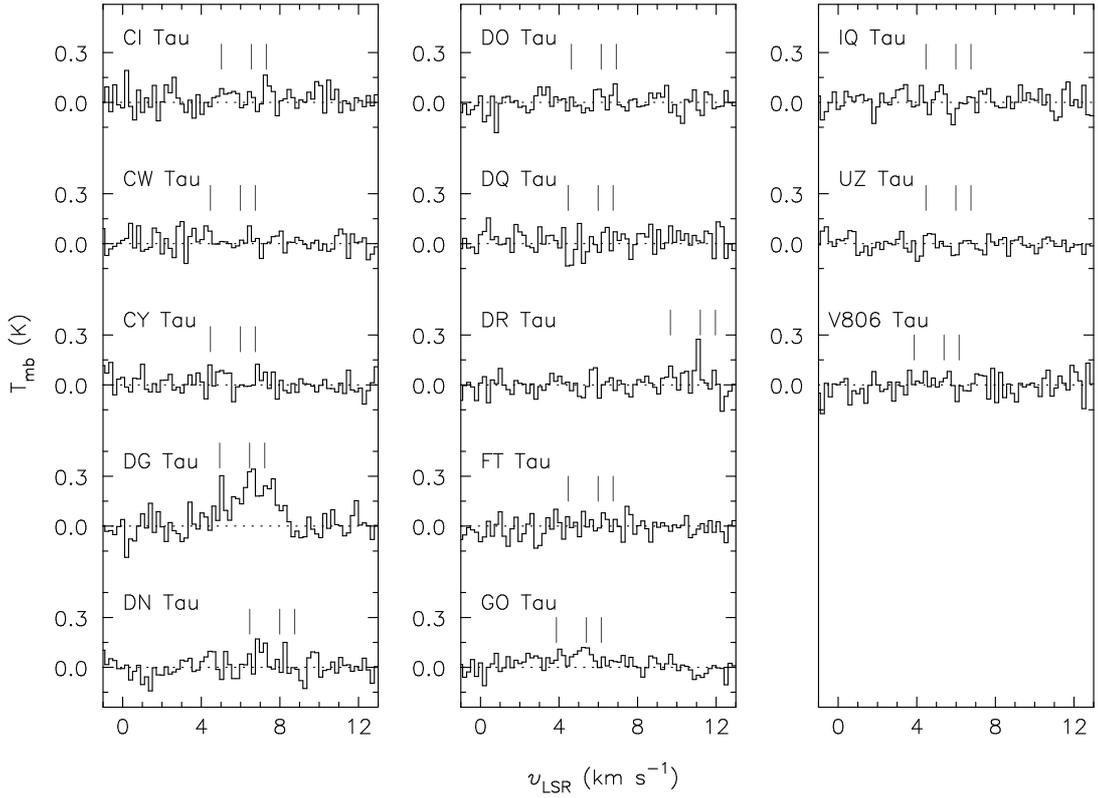} 
\caption{CN\,($J$\,=\,2--1) hyperfine line observations of each of our 13 disks in Taurus. The three vertical lines above each spectrum indicate the expected positions of the three strongest hyperfine lines, located at 226874.1660, 226874.7450, and 226875.8970\,GHz. The spectra have their velocity axis defined such that the strongest hyperfine component, at 226.8747450\,GHz,
should appear at the source velocity ($\upsilon_{\rm LSR}$). In cases where the source velocity is unknown, the Taurus cloud velocity of 6.0\,km\,s$^{-1}$ is indicated instead, as a reference. We report CN detections toward CI\,Tau, DG\,Tau, DR\,Tau, and GO\,Tau only. \label{cnfig} \vspace{1.5cm} } 
\end{figure*}

The data were calibrated by the RxA pipeline and reduced with the Starlink KAPPA software. The edges of each spectrum were removed and a low-order ($p$\,=\,2) polynomial baseline was subtracted. Multiple integrations on individual lines were averaged at this stage. The resulting spectrum was smoothed using a moving square `averaging' box of width 3 channels. We then binned the smoothed data by 5 channels. The final data resolution is 152.5\,kHz (or 0.17\,km\,s$^{-1}$ for HCO$^{+}$ and HCN, and 0.20\,km\,s$^{-1}$ for CN). Finally, the reduced spectra were converted from $T_{A}^{*}$ to main beam temperatures $T_{\rm mb}$ using a beam efficiency factor of $\eta$\,=\,0.65.\footnote{http://docs.jach.hawaii.edu/JCMT/HET/GUIDE/het\_guide/} The final spectra are shown in Figs.~\ref{hcofig}--\ref{cnfig}, and the resulting rms noise levels are reported in Tables \ref{hcohcnlines} and \ref{cnlines}. The analysis of the reduced data was performed using the GILDAS CLASS software to take advantage of the available hyperfine line-fitting algorithms for blended lines. Results of the Gaussian fits to the data, including the peak temperature ($T_{\rm mb}$), line widths (FWHM), and the integrated line intensities ($\int $$T_{\rm mb}$\,$\delta \nu$), are also reported for HCO$^{+}$ and HCN in Table~\ref{hcohcnlines} and for the CN hyperfine lines in Table~\ref{cnlines}.


\begin{figure*} 
\centering 
\includegraphics[width=18cm]{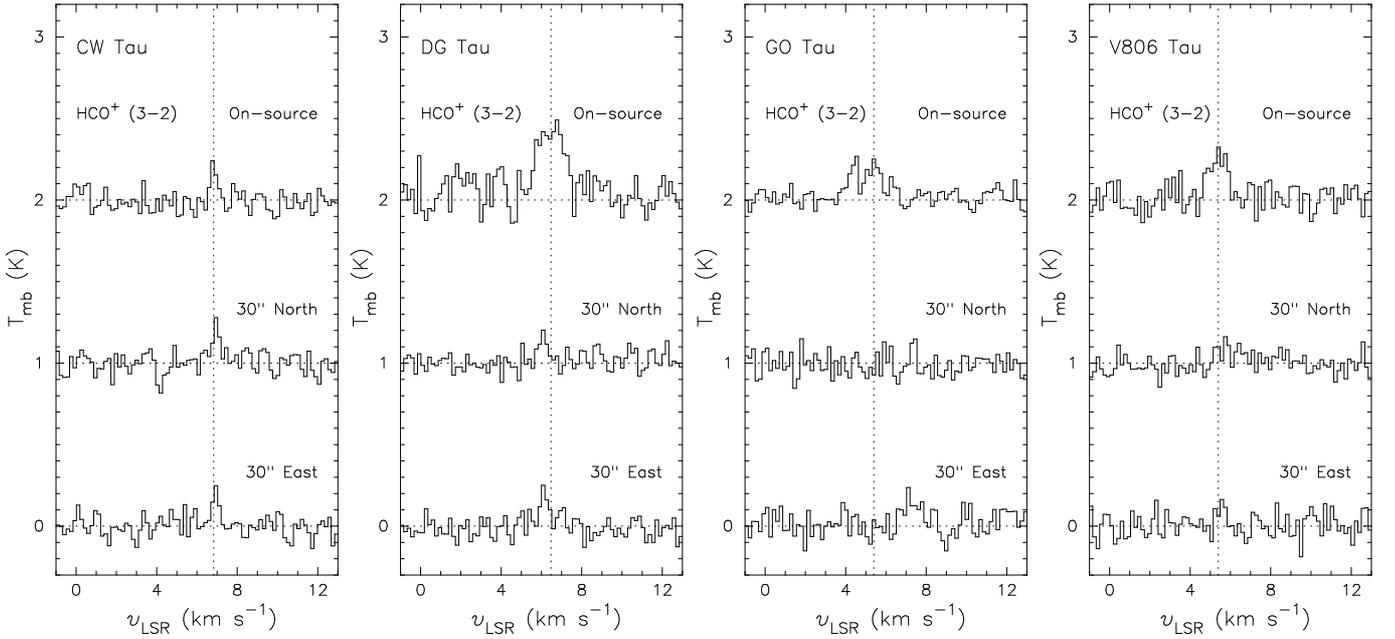} 
\caption{From left to right, the HCO$^{+}$\,($J$\,=\,3--2) on-source and 30$''$ offset observations for CW Tau, DG Tau, GO Tau, and V806 Tau. These 4 sources show the strongest HCO$^{+}$ emission profiles in our subsample. A dotted line marks the central radial velocity $\upsilon_{\rm LSR}$ in km\,s$^{-1}$ for a Gaussian fit to the HCO$^{+}$ on-source emission. In all cases, HCO$^{+}$ emission is detected at the off positions. However, with the exception of CW Tau, it is always at a separate velocity or a weaker intensity, which is why we attribute the emission toward CW Tau to the surrounding cloud and \emph{not} the disk. \label{hcodetections} } 
\end{figure*} 
 

\begin{table*}[!ht]
\begin{minipage}[t]{\linewidth}
\caption{HCO$^{+}$ and HCN line measurements. \label{hcohcnlines} }
\centering
\begin{tabular}{l|cccccccc|cc}
\hline
\hline
\noalign{\smallskip}

Source & \multicolumn{8}{c}{HCO$^{+}$\,(3--2)} ~ & \multicolumn{2}{|c}{HCN\,(3--2) } \\ 

~ & $\upsilon_{\rm LSR}$ & $\sigma_{\rm rms}$ & max $T_{\rm mb}$ & S/N & peak $T_{\rm mb}$ & FWHM & $\Sigma$\,$T_{\rm mb}$\,$\Delta \nu$ & $\int  $$T_{\rm mb}$\,$\delta \nu$ & $\sigma_{\rm rms}$ & $\int T_{\rm mb}$\,$\delta \nu$ \\ 

~ & [km\,s$^{-1}$] & [mK] & [mK]$^{a}$ & ~ & [mK]$^{b}$ & [km\,s$^{-1}$] & [mK\,km\,s$^{-1}$]$^{c}$ & [mK\,km\,s$^{-1}$]$^{d}$ & [mK] & [mK\,km\,s$^{-1}$]$^{e}$ \\

\hline
\noalign{\smallskip}

CI Tau & 6.56 & 57 & 181 & 3.2 & 161 & 0.68 & 105 & 117\,$\pm$\,32 & 76 & $<$\,112 \\
CW Tau & 6.83\,$^{f}$ & 55 & 241 & 4.4 & 259 & 0.39 & 92 & 108\,$\pm$\,22 & 76 & $<$\,112 \\
~~~off 30$''$\,N & 6.95 & 61 & 279 & 4.5 & 270 & 0.38 & 95 & 109\,$\pm$\,27 & $\cdots$ & $\cdots$ \\
~~~off 30$''$\,E & 6.93 & 61 & 248 & 4.1 & 251 & 0.38 & 88 & 101\,$\pm$\,24 & $\cdots$ & $\cdots$ \\
CY Tau & - & 62 & - & - & - & 1.00 & - & $<$\,92 & 110 & $<$\,163 \\
DG Tau & 6.47 & 100 & 491 & 4.9 & 424 & 1.44 & 695 & 652\,$\pm$\,72 & 41 & $<$\,61 \\
~~~off 30$''$\,N & 6.07 & 58 & 203 & 3.5 & 174 & 0.43 & 89 & 79\,$\pm$\,23 & $\cdots$ & $\cdots$ \\
~~~off 30$''$\,E & 6.13 & 65 & 251 & 3.9 & 247 & 0.47 & 107 & 124\,$\pm$\,29 & $\cdots$ & $\cdots$ \\
DN Tau & 8.00 & 48 & 122 & 2.6 & 128 & 0.44 & 56 & 60\,$\pm$\,18 & 85 & $<$\,126 \\
DO Tau  & 6.16 & 48 & 182 & 3.8 & 127 & 0.86 & 115 & 116\,$\pm$\,26 & 67 & $<$\,99 \\
DQ Tau & - & 59 & - & - & - & 1.00 & - & $<$\,88 & 81 & $<$\,120 \\
DR Tau & - & 65 & - & - & - & 1.00 & - & $<$\,97 & 93 & $<$\,138 \\
FT Tau & - & 54 & - & - & - & 1.00 & - & $<$\,80 & 87 & $<$\,129 \\
GO Tau & 5.05 & 47 & 251 & 5.3 & 186\,$^{g}$ & 1.93 & 327 & 383\,$\pm$\,39 & 71 & $<$\,105 \\
~~~off 30$''$\,N & 7.37 & 68 & 148 & 2.2 & 279 & 0.17 & 46 & 51\,$\pm$\,19 & $\cdots$ & $\cdots$ \\
~~~off 30$''$\,E & 7.30 & 71 & 126 & 1.8 & 155 & 1.03 & 172 & 170\,$\pm$\,44 & $\cdots$ & $\cdots$ \\
IQ Tau & - & 80 & - & - & - & 1.00 & - & $<$\,119 & 91 & $<$\,135 \\
UZ Tau & - & 66 & - & - & - & 1.00 & - & $<$\,98 & 80 & $<$\,119\\
V806 Tau & 5.40 & 77 & 345 & 4.2 & 271 & 0.97 & 297 & 279\,$\pm$\,45 & 100 & $<$\,148 \\
~~~off 30$''$\,N & 5.80 & 61 & 162 & 2.6 & 201 & 0.25 & 82 & 52\,$\pm$\,20 & $\cdots$ & $\cdots$ \\
~~~off 30$''$\,E & 5.59 & 68 & 164 & 2.4 & 149 & 0.39 & 76 & 62\,$\pm$\,28 & $\cdots$ & $\cdots$ \\

\hline
\noalign{\smallskip}

GO Tau 1$^{h}$ & 4.40 & 47 & 268 & 5.7 & 237 & 0.54 & 100 & 136\,$\pm$\,28 & $\cdots$ & $\cdots$ \\
GO Tau 2 & 5.40 & 47 & 251 & 5.3 & 232 & 0.75 & 179 & 185\,$\pm$\,28 & $\cdots$ & $\cdots$ \\
GO Tau 3 & 6.34 & 47 & 143 & 3.0 & 125 & 0.35 & 50 & 46\,$\pm$\,19 & $\cdots$ & $\cdots$ \\

\hline
\noalign{\smallskip}
\end{tabular}
\end{minipage}

\hfill

\begin{minipage}[h]{\linewidth}
\begin{tiny}
\textbf{Footnotes.} (a) The maximum temperature value in the binned data; (b) The maximum determined by a Gaussian fit to the line using the CLASS Gauss fitting routine; (c) A sum of the individual channels from the binned data to be compared to the integrated fit; (d) For non-detections we report a 3$\sigma$ upper limit where $\sigma$\,=\,1.2\,$\sigma_{\rm rms}$$\sqrt{\Delta V \delta \nu}$ and we assume a line width $\Delta V$ of 1.0\,km\,s$^{-1}$; (e) Upper limits are reported in the same way as for HCO$^{+}$; (f) We have determined that the emission detected here is not associated with the circumstellar disk; (g) Statistics are reported for a single Gaussian fit, which may not be realistic for this source; (h) Statistics for the individual peaks toward GO Tau, measured as they appear from left to right in Fig.~\ref{hcofig}. 

\end{tiny}
\hfill
\end{minipage} 
\end{table*} 



\begin{table*}[th!] 
\begin{minipage}[t]{\linewidth} 
\caption{CN line measurements. \label{cnlines} } 
\centering 
\begin{tabular}{l|ccc|ccc|ccc|cc} 
\hline 
\hline 
\noalign{\smallskip} 

Source & \multicolumn{3}{c|}{CN\,(2\,$_{3/2, 3/2}$--1\,$_{5/2, 5/2}$)} & \multicolumn{3}{c|}{CN\,(2\,$_{3/2, 5/2}$--1\,$_{5/2, 7/2}$)} & \multicolumn{3}{c|}{CN\,(2\,$_{3/2, 1/2}$--1\,$_{5/2, 3/2}$)} & \multicolumn{2}{c}{Average Stats}  \\ 

~ & $\upsilon_{\rm LSR}$ & Peak & $\int $$T_{\rm mb}$\,$\delta \nu$ & $\upsilon_{\rm LSR}$ & Peak & $\int $$T_{\rm mb}$\,$\delta \nu$ & $\upsilon_{\rm LSR}$ & Peak & $\int $$T_{\rm mb}$\,$\delta \nu$ & FWHM & $\sigma_{\rm rms}$ \\ 

~ & [km\,s$^{-1}$] & [mK] & [mK\,km\,s$^{-1}$] & [km\,s$^{-1}$] & [mK] & [mK\,km\,s$^{-1}$] & [km\,s$^{-1}$] & [mK] & [mK\,km\,s$^{-1}$] & [km\,s$^{-1}$] & [mK] \\ 

\hline
\noalign{\smallskip}

CI Tau & 7.33 & 230 & 52\,$\pm$\,19 & 6.51 & 76 & 16\,$\pm$\,19 & 5.28 & 65 & 58\,$\pm$\,35 & 0.42 & 63 \\
CW Tau & - & - & $<$\,94 & - & - & $<$\,94 & - & - & $<$\,94 & 1.00 & 58 \\
CY Tau & - & - & $<$\,97 & - & - & $<$\,97 & - & - & $<$\,97 & 1.00 & 60 \\
DG Tau & 7.63 & 198 & 171\,$\pm$\,94 & 6.45 & 279 & 414\,$\pm$\,124 & 5.03 & 294 & 76\,$\pm$\,30 & 0.82 & 63 \\
DN Tau & - & - & $<$\,104 & - & - & $<$\,104 & - & - & $<$\,104 & 1.00 & 64 \\
DO Tau & - & - & $<$\,90 & - & - & $<$\,90 & - & - & $<$\,90 & 1.00 & 56 \\
DQ Tau & - & - & $<$\,98 & - & - & $<$\,98 & - & - & $<$\,98 & 1.00 & 61 \\
DR Tau & 9.65 & 119 & 56\,$\pm$\,26 & 11.06 & 297 & 87\,$\pm$\,20 & 11.93 & 110 & 24\,$\pm$\,16 & 0.31 & 55 \\
FT Tau & - & - & $<$\,88 & - & - & $<$\,88 & - & - & $<$\,88 & 1.00 & 55 \\
GO Tau & 2.56 & 55 & 101\,$\pm$\,41 & 4.07 & 103 & 37\,$\pm$\,19 & 5.24 & 116 & 129\,$\pm$\,32 & 1.03 & 42 \\
IQ Tau & - & - & $<$\,91 & - & - & $<$\,91 & - & - & $<$\,91 & 1.00 & 57 \\
UZ Tau & - & - & $<$\,79 & - & - & $<$\,79 & - & - & $<$\,79 & 1.00 & 49 \\
V806 Tau & - & - & $<$\,102 & - & - & $<$\,102 & - & - & $<$\,102 & 1.00 & 63 \\

\hline
\noalign{\smallskip}
\end{tabular}
\end{minipage}

\hfill

\begin{minipage}[h]{\linewidth}
\begin{tiny}
\textbf{Notes.} The velocities reported here are for a $\upsilon_{\rm LSR}$\,=\,0 centered at the theoretical frequency for the strongest hyperfine component, at 226.8747450\,GHz. All reported peaks were determined from a Gaussian fit. All upper limits are given by 3\,$\times$\,1.2\,$\sigma_{\rm rms}$\,$\sqrt{\Delta V \delta \nu }$ following \citet{jorgensen2004} where $\Delta V$ is set to 1.0\,km\,s$^{-1}$.

\end{tiny}
\hfill
\end{minipage}
\end{table*}

\section{Results}
\label{results}

\begin{table*}[ht!]
\caption{Molecular line fluxes. \label{fluxdata}}
\begin{minipage}[h]{\linewidth}
\centering
\begin{tabular}{lcccccccccccc}
\hline
\hline
\noalign{\smallskip}

(1)        & (2)	     & (3)	   & (4)  & (5) 	    & (6)	 & (7)  & (8)  	           & (9)	& (10) & (11)		  & (12)       & (13)  \\
Source     & F$_{\rm ^{13}CO}$  & $^{13}$CO   & Ref. & F$_{\rm HCO^+}$ & HCO$^+$	 & Ref. & F$_{\rm HCN}$	   & HCN	& Ref. & F$_{\rm CN}$	  & CN         & Ref.  \\
Name       & [Jy\,km\,s$^{-1}$] & Trans.      & ~    & [Jy\,km\,s$^{-1}$] 	    & Trans.	 & ~	& [Jy\,km\,s$^{-1}$]  	   & Trans.	& ~    & [Jy\,km\,s$^{-1}$]		  & Trans.     & ~     \\
\hline
AA Tau     & 12.4	     & 2--1	   & 2   & 1.9  	    & 1--0	 & 8	& 1.8		   & 1--0	& 8    & 3.8		  & 1--0       & 8     \\
CI Tau     & -  	     & -	   & -   & 2.4\,$\pm$\,0.7  & 3--2	 & 1	& $<$\,1.6	   & 3--2	& 1    & 2.6\,$\pm$\,1.5  & 2--1       & 1     \\ 
CW Tau     & 8.3	     & 2--1	   & 2   & $<$\,1.8         & 3--2	 & 1	& $<$\,1.6	   & 3--2	& 1    & $<$\,1.9	  & 2--1       & 1     \\
CY Tau     & -  	     & -	   & -   & $<$\,1.9	    & 3--2	 & 1	& $<$\,2.3	   & 3--2	& 1    & $<$\,2.0	  & 2--1       & 1     \\
DG Tau     & 40.0	     & 1--0	   & 3   & 13.5\,$\pm$\,1.5 & 3--2	 & 1	& $<$\,0.8	   & 3--2	& 1    & 13.7\,$\pm$\,5.1 & 2--1       & 1     \\
DL Tau     & $<$\,2.1	     & 3--2	   & 2   & -		    & - 	 & -	& -		   & -  	& -    & -		  & -	       & -     \\
DM Tau     & 6.5	     & 2--1	   & 4   & 4.1\,$\pm$\,0.5  & 3--2	 & 5	& 2.0\,$\pm$\,0.3  & 1--0	& 6    & 8.7\,$\pm$\,0.4  & 2--1       & 6     \\
DN Tau     & 14.5	     & 2--1	   & 2   & 1.2\,$\pm$\,0.4  & 3--2	 & 1	& $<$\,1.8	   & 3--2	& 1    & $<$\,2.2	  & 2--1       & 1     \\
DO Tau     & 31.0	     & 2--1	   & 2   & 2.4\,$\pm$\,0.5  & 3--2	 & 1	& $<$\,1.4	   & 3--2	& 1    & $<$\,1.9	  & 2--1       & 1     \\
DQ Tau     & -  	     & -	   & -   & $<$\,1.8	    & 3--2	 & 1	& $<$\,1.7	   & 3--2	& 1    & $<$\,2.0	  & 2--1       & 1     \\
DR Tau     & 4.3\,$\pm$\,1.2 & 3--2	   & 5   & 3.7\,$\pm$\,0.4  & 4--3	 & 9	& $<$\,1.9	   & 3--2	& 1    & 3.5\,$\pm$\,1.3  & 2--1       & 1     \\
FT Tau     & -  	     & -	   & -   & $<$\,1.7	    & 3--2	 & 1	& $<$\,1.8	   & 3--2	& 1    & $<$\,1.8	  & 2--1       & 1     \\
GG Tau     & 5.8\,$\pm$\,0.2 & 2--1	   & 6   & 6.7\,$\pm$\,0.9  & 3--2	 & 6	& 2.8\,$\pm$\,0.7  & 3--2	& 6    & 8.4\,$\pm$\,0.4  & 2--1       & 6     \\
GM Aur     & 9.3\,$\pm$\,2.7 & 3--2	   & 5   & 9.9\,$\pm$\,1.2  & 4--3	 & 9	& -		   & -  	& -    & -		  & -	       & -     \\
GO Tau     & 4.1\,$\pm$\,1.2 & 3--2	   & 5   & 7.9\,$\pm$\,0.8  & 3--2	 & 1	& $<$\,1.5	   & 3--2	& 1    & 5.5\,$\pm$\,1.9  & 2--1       & 1     \\
IQ Tau     & -  	     & -	   & -   & $<$\,2.5	    & 3--2	 & 1	& $<$\,1.9	   & 3--2	& 1    & $<$\,1.9	  & 2--1       & 1     \\
RY Tau     & 4.3\,$\pm$\,1.2 & 3--2	   & 5   & 1.7\,$\pm$\,0.7  & 4--3	 & 9	& 1.6		   & 1--0	& 8    & 0.8		  & 1--0       & 8     \\
T Tau 	   & 3.0\,$\pm$\,0.2 & 2--1	   & 7   & 3.1\,$\pm$\,0.1  & 1--0	 & 7	& 1.9\,$\pm$\,0.8  & 2--1	& 10   & -		  & -	       & -     \\
UZ Tau     & -  	     & -	   & -   & $<$\,2.0	    & 3--2	 & 1	& $<$\,1.7	   & 3--2	& 1    & $<$\,1.6	  & 2--1       & 1     \\
V806 Tau   & -  	     & -	   & -   & 5.8\,$\pm$\,0.9  & 3--2	 & 1	& $<$\,2.1	   & 3--2	& 1    & $<$\,2.1	  & 2--1       & 1     \\
V892 Tau   & -  	     & -	   & -   & -		    & - 	 & -	& -		   & -  	& -    & -		  & -	       & -     \\

\hline
\noalign{\smallskip}
\end{tabular}
\end{minipage}
\hfill
\begin{minipage}[]{\linewidth}
\textbf{References.} 1.~This work; 2.~\citet{greaves2005}; 3.~\citet{kitamura1996b}; 4.~\citet{panic2008}; 5.~\citet{thi2001}; 6.~\citet{dutrey1997}; 7.~\citet{hogerheijde1998}; 8.~\citet{kesslerphd2004}; 9.~\citet{greaves2004}; and 10.~\citet{yun1999}. 
\end{minipage}

\end{table*}

In Fig.~\ref{hcofig} we see that only 7 of our 13 sources show detected emission lines of HCO$^{+}$\,($J$\,=\,3--2)\,: CI Tau, CW Tau, DG Tau, DN Tau, DO Tau, GO Tau, and V806 Tau. The detected lines have intensities of $T_{\rm mb}$\,=\,122--491\,mK and are well fit by single Gaussian profiles with widths of $\sim$1\,km\,s$^{-1}$\,FWHM. Remarkably, no strong double-peaked line profiles characteristic of inclined disks are found. Only DG Tau and GO Tau show wider lines of 1.44 and 1.93\,km\,s$^{-1}$, respectively. These sources also represent two of the strongest lines, with respective peaks $T_{\rm mb}$ of 491 and 251\,mK. GO\,Tau is the only source for which the line profile is not well described by a single Gaussian. Instead, it consists of a superposition of three, narrow ($\sim$0.5\,km\,s$^{-1}$) components. \citet{greaves2004} and \citet{thi2004} also comment on the complex line profiles of GO Tau, attributing some of the components to surrounding cloud material. Sect.~\ref{others} discusses the literature and these detections further.

Fig.~\ref{hcodetections} shows the HCO$^+$ spectra at $30''$ east and $30''$ north of the four sources with the brightest detected HCO$^+$ lines: CW Tau, DG Tau, GO Tau, and V806 Tau. In all four cases, hints of extended cloud emission are seen. However, only toward CW Tau are they of comparable strength and shape as the on-source emission, and we no longer consider this a detection of emission from the disk. This brings our total \emph{disk} detections in HCO$^{+}$ down to 6. For the other sources, the off-source emission is $<$\,30\% of the on-source line strength, which we determine is sufficiently small to be inconsequential to our trends analysis (Sect.~\ref{trends}); while our more detailed modeling of DG Tau considers the contributions that remnant envelopes and other circumstellar material can add to the line emission (Sect.~\ref{modeling}). Finally, because the remaining HCO$^+$ detections (CI Tau, DN Tau, and DO Tau) are without accompanying off-source measurements and they exhibit relatively narrow lines, their disk origin must also be confirmed. In the meantime, because their line widths are broader than what is typically expected from extended cloud material ($\sim$0.3\,km\,s$^{-1}$ for a cloud with a $T$ of 25\,K and an H$_2$ number density of 10$^4$\,cm$^{-3}$), we report these sources as detected, in following suit with previous single-dish studies. However, our results highlight the relevance and necessity of performing off-source measurements during single-dish surveys, even for molecular species that preferentially trace much denser material.

No HCN\,(3--2) emission was detected toward any of our targets, down to 3$\sigma$ upper limits of 41--110\,mK in 0.17\,km\,s$^{-1}$ bins (see Table \ref{hcohcnlines}). Following the recipe of \citet{jorgensen2004}, this corresponds to upper limits on the integrated intensities ranging from 61 to 163\,mK\,km~s$^{-1}$. Clearly, toward the sources in our Taurus subsample, the HCN lines are at least a factor of 3 weaker than the HCO$^+$ lines (and a factor of 12 weaker toward DG Tau where a much deeper integration resulted in a lower noise level). 

Two sources, DG Tau and GO Tau, show clear detections of the CN\,(2--1) triplet at frequencies of 226.8741660, 226.8747450, and 226.8758970\,GHz.\footnote{The JPL Molecular Spectroscopy Catalog http://spec.jpl.nasa.gov} Two additional sources, CI Tau and DR Tau, show tentative detections. No emission was detected from the other targets where the typical rms noise level was 57\,mK in 0.20\,km\,s$^{-1}$ bins. Two suggestive peaks in the DO Tau spectrum at the correct position for the CN hyperfine components are likely noise.


\begin{figure*}[tbh!] 
\centering 
\includegraphics[width=16cm]{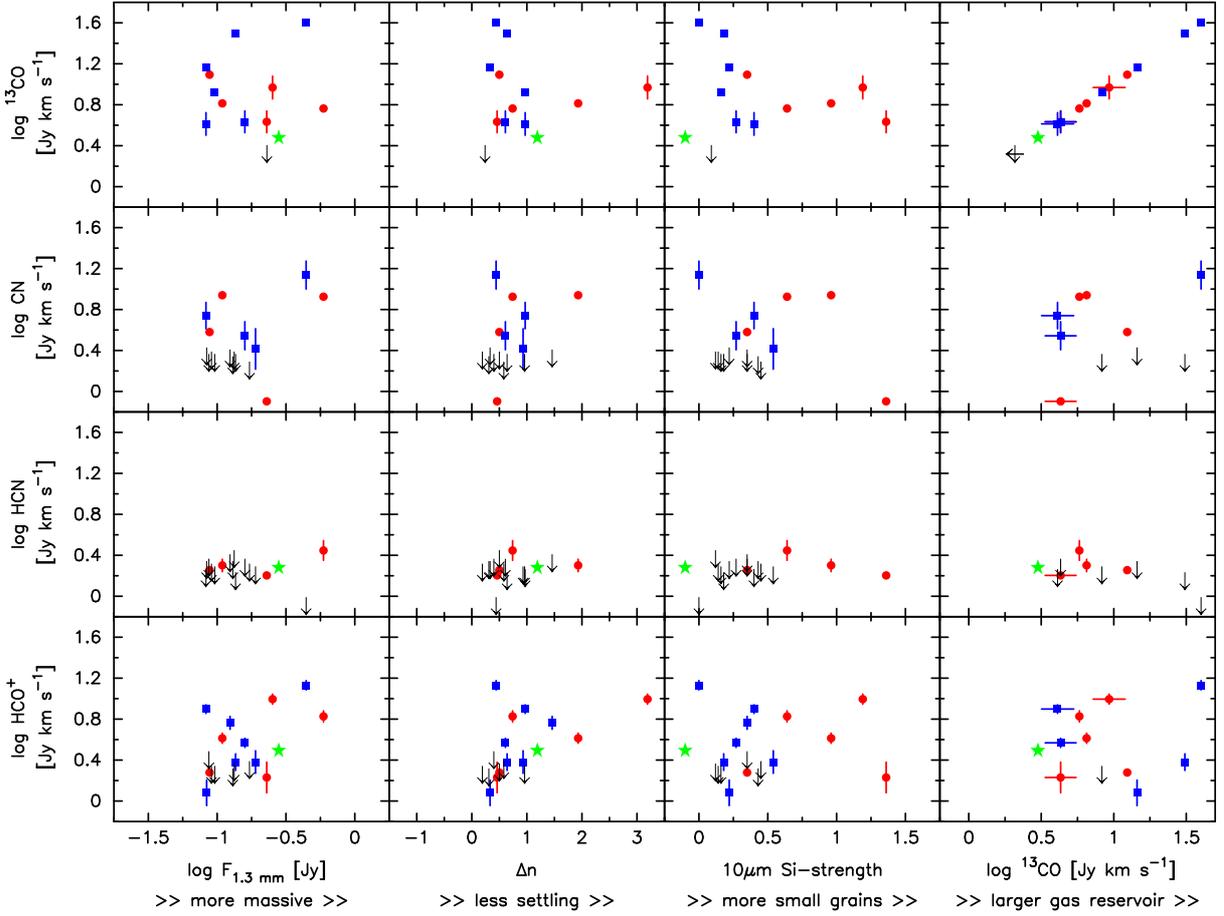} 
\caption{A log plot of the integrated line intensities in Jy\,km\,s$^{-1}$ versus several disk parameters that probe the total dust mass ($F_{\rm 1.3\,mm}$), dust settling via the mid-infrared spectral slope ($\Delta n$), amount of small grains via the 10$\rm \mu$m silicate emission feature strength (Si-strength), and the total gas mass (inferred from the $^{13}$CO line flux). Blue squares represent our JCMT observations, red circles represent data from the literature, and black arrows are used to indicate all upper limits. A green star symbol represents T~Tau, whose line emission is contaminated by a large remnant envelope in the single-dish data plotted. The upper-right panel illustrates the dynamic range of the available gas reservoirs. (Color plots are available online) \label{trends1} }  
\end{figure*} 


\section{Trends}
\label{trends}

The previous section shows 6 detections of HCO$^+$ emission, no detections of HCN emission, and 4 detections of CN emission from our subsample of 13 observed sources. Looking at the complete sample of 21 sources yields the following statistics for detections of HCO$^+$\,(14/21), HCN\,(5/21), and CN\,(8/21) toward the T~Tauri disk population thus far in Taurus. Half of the reported detections are taken from the literature (see Table \ref{fluxdata}) and these often include rotational transitions other than the ones observed by us. However, in the present section we focus on general trends in the detected emission strengths, and we assume that the different transitions correspond to line strength differences of no more than a factor of $\sim$2.

Our goal is to investigate whether the line strengths (and the upper limits) from the Taurus sample show any correlation with disk or stellar properties. For this purpose, we converted the line strengths to integrated line fluxes (in Jy\,km\,s$^{-1}$) in order to remove the dependences on beam size. We used the formula:

\begin{equation}
S_{\nu} = \frac{\Omega^2 [\rm{''}]}{\lambda^2 [\rm{cm}]} \cdot 7.354 \times 10^{-4} \cdot T_{\rm MB} [\rm{Jy/K}]
\end{equation}

\noindent where $\Omega$ is the primary beam in arc seconds, $\lambda$ is the observed wavelength in cm, and $T_{\rm MB}$ is the observed main beam temperature in K. The assembled line fluxes for the entire sample are provided in Table \ref{fluxdata}. To trace the disk properties we use the millimeter continuum flux $F_{\rm 1.3\,mm}$, the difference in the mid-infrared slope $\Delta n$ as defined by \citet{furlan2006}, and the 10-$\rm \mu$m silicate feature strength, also from \citet{furlan2006}. Respectively, these trace the disk (dust) mass, dust settling, and small-dust content. Stellar properties are probed by stellar mass M$_\star$, H$\alpha$ equivalent width, bolometric luminosity $L_{\rm bol}$, and X-ray luminosity $L_{\rm X}$. We stress that all quantities are \emph{observed} quantities, and that a relation (or lack thereof) between the HCO$^+$ line strength and the 1.3\,mm continuum flux, for example, does not immediately imply a connection (or lack thereof) between HCO$^+$ abundance and disk (dust) mass. To draw the latter conclusion, detailed modeling of the emission mechanisms is required, which is the topic of Sect.~\ref{modeling}.


\begin{figure*}[tbh!] 
\centering 
\includegraphics[width=17cm]{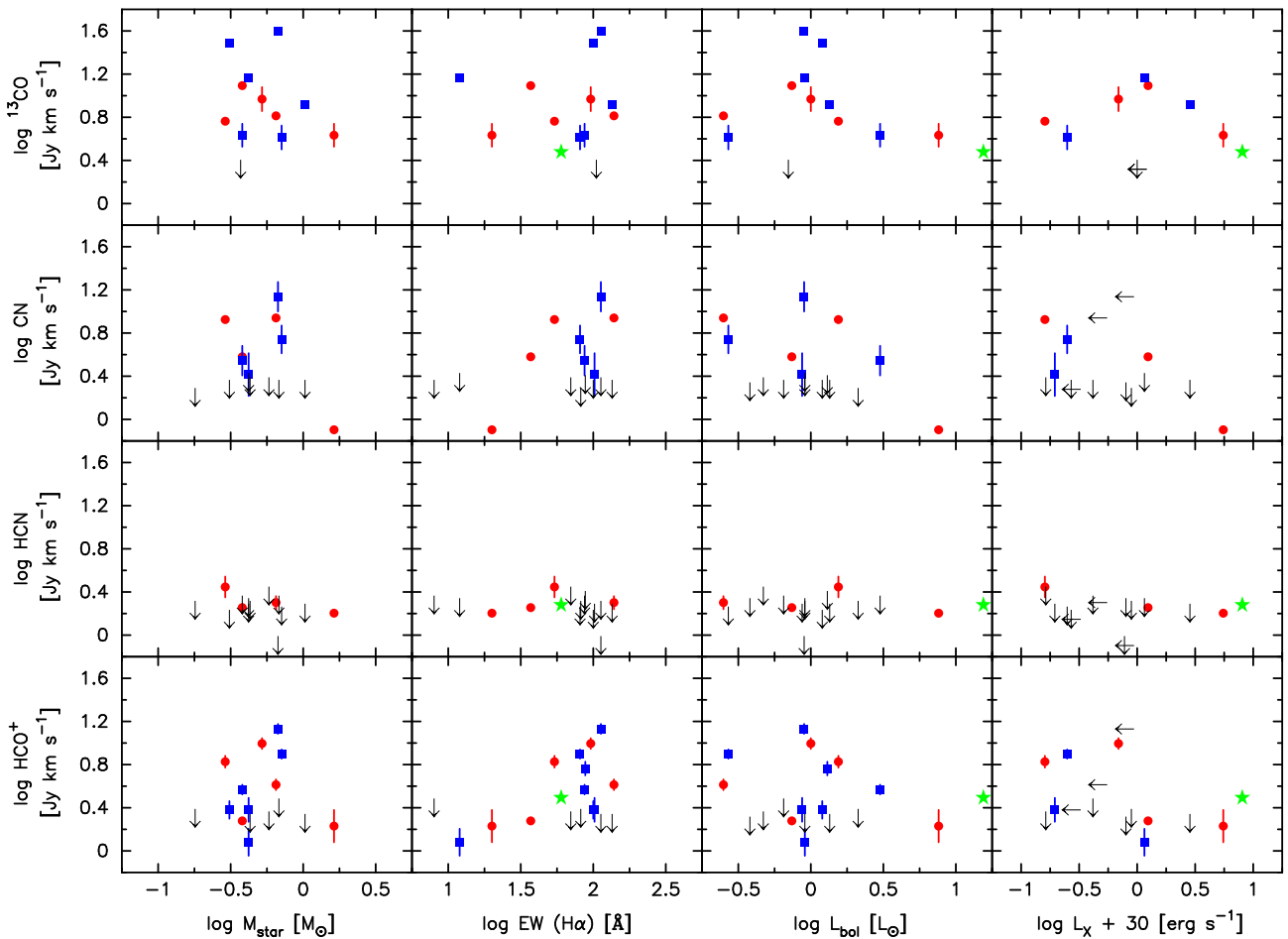} 
\caption{A plot of the integrated line intensities in Jy\,km\,s$^{-1}$ versus several stellar parameters from the literature, including: the stellar mass (M$_{\star}$), the H$\alpha$ equivalent width (an accretion tracer), the bolometric luminosity ($L_{\rm bol}$), and the X-ray luminosity ($L_{\rm X}$). See Fig.~\ref{trends1} for an explanation of the symbols. (Color plots are available online) \label{trends4} } 
\end{figure*} 


One of the guiding questions in our research was, whether disks with a lower dust content or with more dust settling have larger gas volumes subject to UV irradiation, and therefore show comparatively stronger lines of HCO$^+$ (more ionization) and CN (more photodissociation). In the following figures, we will explicitely investigate these relations. Fig.~\ref{trends1} plots the observed HCO$^+$, HCN, CN, and $^{13}$CO line fluxes against $F_{\rm 1.3\,mm}$, $\Delta n$, 10-$\mu$m silicate feature strength, and $^{13}$CO line flux. The latter quantity serves as a tracer of the total molecular gas content. We choose to plot the logarithm of these quantities because we are primarily interested in large-scale trends and because we want to suppress differences of a factor of $\sim$2 due to variations in source distance, inclination, luminosity, and plotted line transitions. No immediate trends are present in the panels of Fig.~\ref{trends1}. Instead, we find that: (1) the millimeter continuum flux alone is not a clear predictor of line strength since both lower- and higher-mass disks exhibit detections and non-detections alike in all species; (2) disks with and without significant dust settling exhibit equally strong line strengths all around; and (3) a larger $^{13}$CO reservoir is not necessarily accompanied by stronger HCO$^+$, HCN, and CN lines. These results hold for a $^{13}$CO line flux spanning a factor $\geq$10 in dynamic range. 

We then considered the same HCO$^+$, HCN, and CN line strengths versus $F_{\rm 1.3mm}$, $\Delta n$, 10-$\mu$m silicate feature strength, and $^{13}$CO line flux, but this time only after normalizing the line fluxes to the $^{13}$CO line flux (not shown). This relation represents the line flux per unit gas mass, if we assume that the $^{13}$CO line fluxes are reliable tracers of the total amount of molecular gas in each disk. No obvious trends were visible there either. Finally, we explored the equivalent plots of line flux, but then normalizing to $F_{\rm 1.3mm}$, or the line flux per unit dust emission. Again, no trends emerged. Together, the results show that there is no clear correlation between the observed line fluxes of HCO$^+$, HCN, or CN, and the disk dust mass, the degree of settling, the amount of small particles in the disk, or the disk gas mass. Echoing our earlier remarks, we emphasize that only detailed modeling of the emission lines can prove the presence or absence of such a correlation; but judging from the figures that plot general trends only, no strong correlation is expected from this sample.

If the molecular line flux does not depend strongly on the properties of the disk, perhaps it depends on the properties of the star. In Fig.~\ref{trends4} we plot the line fluxes and upper limits against stellar mass, H$\alpha$ equivalent width, bolometric luminosity, and X-ray luminosity. Similarly, we also explored the results of normalizing the line fluxes with respect to the $^{13}$CO line flux and $F_{\rm 1.3mm}$. While the UV flux could be expected to depend strongly on the mass accretion rate, no homogeneous set of mass accretion rates is available in the literature for this sample, and these can be expected to be variable as well. Instead, we use the H$\alpha$ equivalent width as a tracer and find no correlation. Indeed, no trends are apparent in any of the plots, suggesting that also such a simple connection between the stellar radiation and the emergent HCO$^+$, HCN, or CN line flux is absent. 

\begin{figure*}[tbh!] 
\centering 
\begin{minipage}[l]{13cm}
\includegraphics[width=13cm]{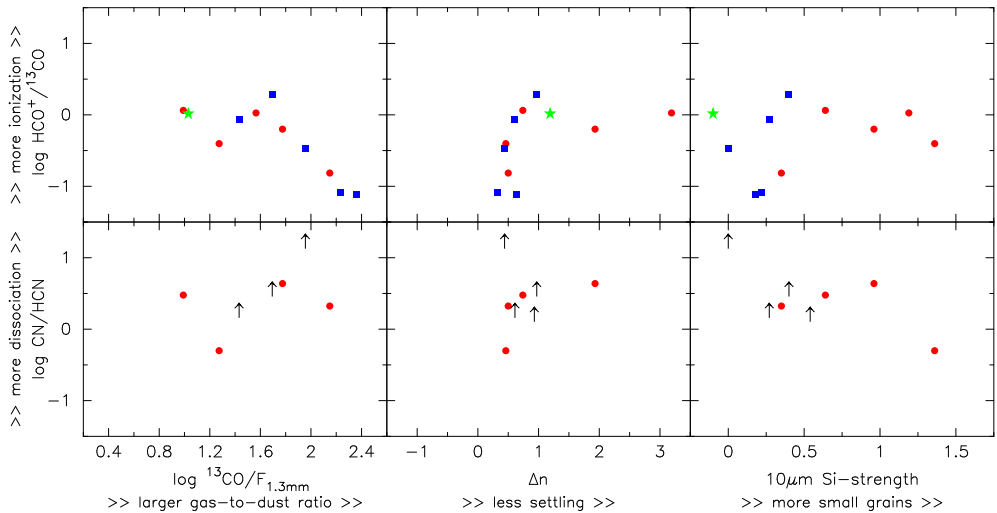}
\end{minipage}
\hfill
\begin{minipage}[r]{5cm}
\caption{Upper panels: The degree of ionization (traced by HCO$^+$/$^{\rm 13}$CO) versus the gas-to-dust ratio ($^{13}$CO/F$_{\rm 1.3mm}$), the difference in the mid-infrared spectral slope between 6, 10, and 25$\mu$m ($\Delta n$), and the strength of the 10$\mu$m silicate emission feature. Lower panels: Similar plots for the photodissociation effect (traced by CN/HCN). See Fig.~\ref{trends1} for an explanation of the symbols. (Color plots are available online) \label{trends7} \vspace{3cm} } 
\end{minipage}
\end{figure*}

In Fig.~\ref{trends7} we plot the specific relations that served as the primary motivation for this study: the ratio of CN over HCN line fluxes and the HCO$^+$/$^{13}$CO flux, respectively tracing the degree of photodissociation and the degree of ionization, versus the gas-to-dust ratio (represented by $^{13}$CO/$F_{\rm 1.3\,mm}$) and the changing mid-infrared slope $\Delta n$, representing dust settling. As was the case for our other plots, no significant correlations are apparent. We note that the decreasing line strength with a larger gas-to-dust ratio in the upper-left panel of Fig.~\ref{trends7}, is the likely effect of the incorporation of the $^{13}$CO factor into each ratio.

In summary, we find \emph{no} correlation between the HCO$^+$, HCN, and CN line fluxes (or their ratios) and any tracer of the disk properties or those of the stellar radiation field. The line flux does not seem to be affected by any of the investigated parameters. We conclude that \emph{the details of the input radiation field}, such as UV and Ly$\alpha$ strengths \citep{bergin2003}, may be the deciding factor in the resulting line fluxes. Other contributing factors include the inner versus outer disk contributions and the temperature and density structure in the line-emitting region, both of which are discussed in Sect.~\ref{modeling}.


\section{Modeling the molecular emission}
\label{modeling}

\subsection{Disk models}

The previous section investigates the observational correlation between the measured HCO$^+$, HCN, and CN line intensities and upper limits, and the disk dust observables, such as millimeter continuum flux and infrared slope. While the emergent line intensity depends on the underlying molecular abundance, other factors including the disk density and temperature structure affect the emerging lines through molecular excitation and line radiative transfer. This section addresses how the modeled abundances of HCO$^+$, HCN, and CN (that can explain the observed emission) are related to the disk dust properties. Rather than developing an \emph{ab initio} description of the disk structure and associated molecular chemistry, this section employs two independent models obtained from the literature \citep{robitaille2006,isella2009} as starting points, and calculates the molecular abundances (assumed constant throughout the disk except in regions of freeze-out) that reproduce our line observations.

In the first modeling approach, we make use of the online SED fitting tool\footnote{Online SED fitting tool, http://caravan.astro.wisc.edu/protostars} of \citet{robitaille2006}, the best-fit parameters of \citet{robitaille2007}, and the visual extinction values from \citet{white2001}.\footnote{No visual extinction was included in \citet{robitaille2007} and therefore our best-fit model may differ slightly from the one listed there.} The continuum radiative transfer code of \citet{whitney2003a} produced the two-dimensional density and temperature structure for the best-fitting model for each source. In some cases (i.e.~CI Tau, DO Tau, DR Tau, and FT Tau), the Robitaille models include remnant envelopes; these are not plotted in the figures but are included in the line calculations. However, since their H$_2$ number densities are $<$\,10$^{5}$\,cm$^{-3}$, the envelopes are not expected to contribute significantly to the HCO$^{+}$, HCN, or CN line emission \citep[e.g.][]{hogerheijde2000a}. This first method relies entirely on spatially \emph{unresolved} continuum data, keeping all other stellar and disk parameters free, even in cases where these properties may be well known.


\begin{figure*}[ht!] 
\centering 
\includegraphics[width=17cm]{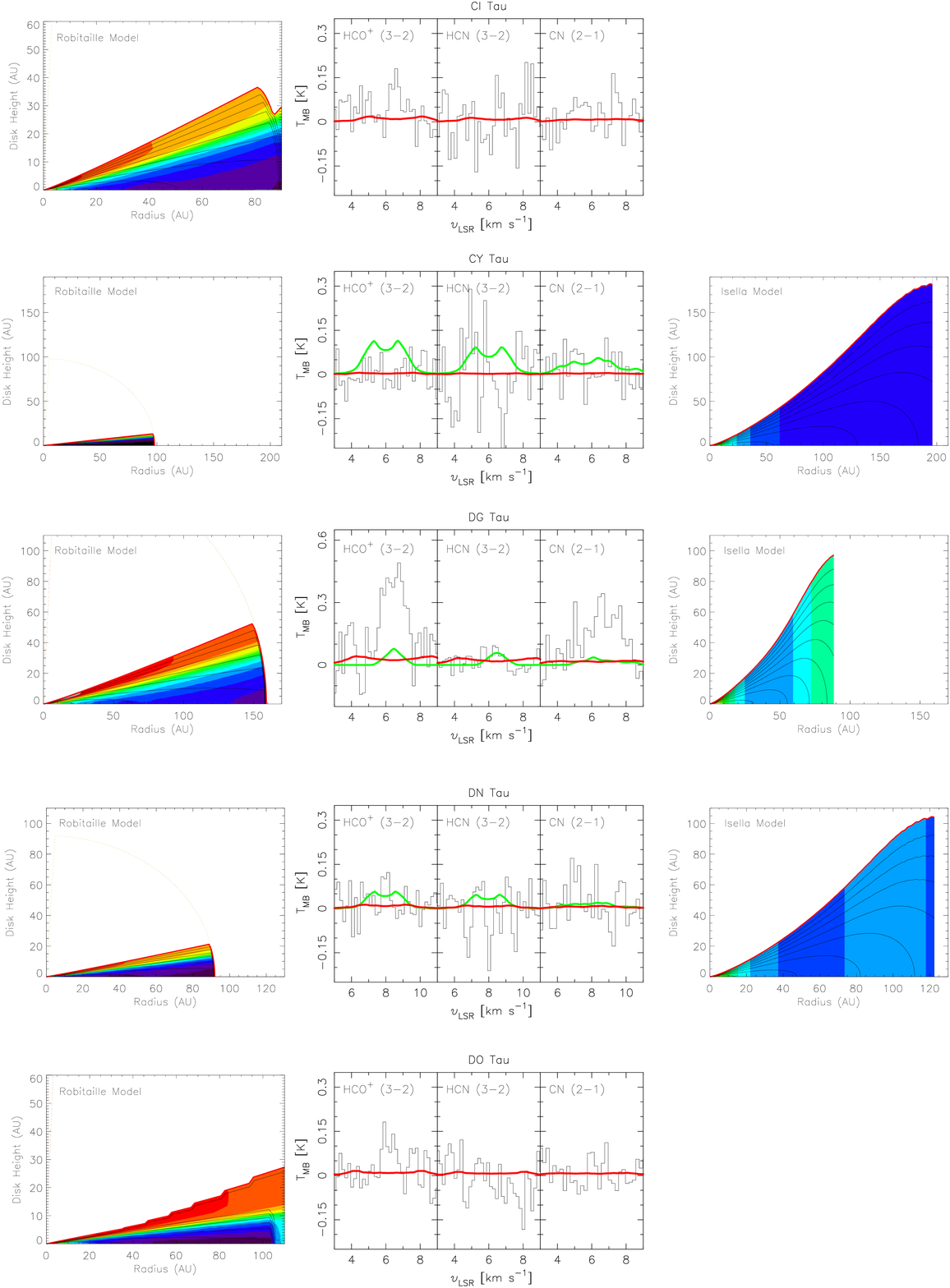} 

\caption{Results of the general analysis. From left to right: the Robitaille disk structure; the line predictions for HCO$^{+}$, HCN, and CN (in red for the Robitaille model, green for the Isella model); and the Isella disk structure (when available). The sources from top to bottom are: CI Tau, CY Tau, DG Tau, DN Tau, and DO Tau. For the disk structures, filled contours correspond to the temperature profile (in K). The temperature levels are identical for all disks, designated at: 10, 15, 20, 25, 30, 35, 40, 45, 50, 55, 60, 75, 100, 150, and 250\,K. A color bar is provided in Fig.~\ref{generalresults2}. The contour lines indicate the log H$_2$ number density (in cm$^{-3}$), indicated at whole number integrals from 3 (disk surface/edge) to 12 (typical midplane density). Models are only pictured for the disks included in the Robitaille and Isella samples. (Color plots are available online) \label{generalresults1} } 

\end{figure*} 

\begin{figure*}[ht!] 
\centering 
\includegraphics[width=16.5cm]{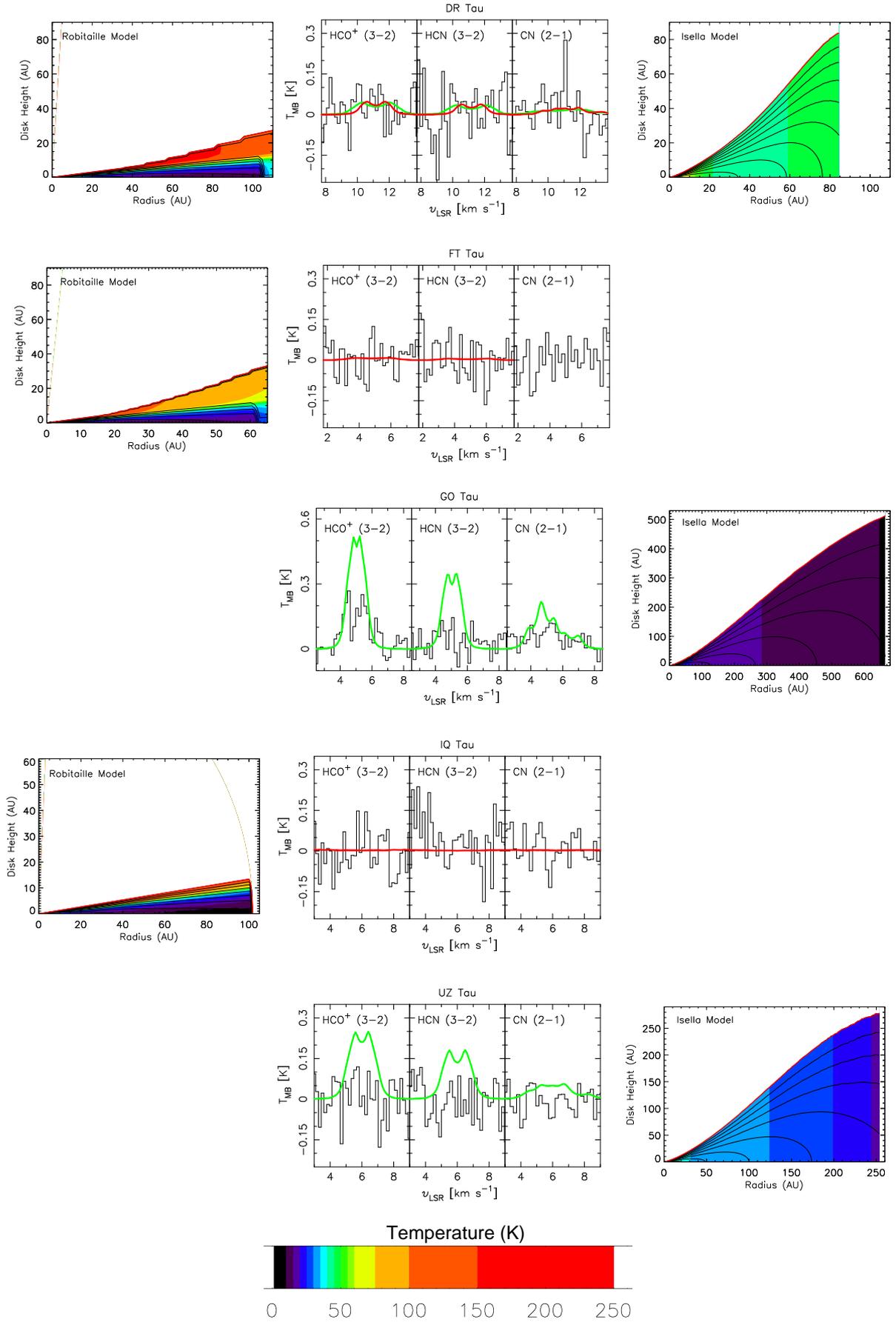} 

\caption{Continued from Fig.~\ref{generalresults1}. The sources from top to bottom are: DR Tau, GO Tau, FT Tau, GO Tau, and IQ Tau. A color bar is provided here for the filled temperature contours and applies to all disks. Models are only pictured for the disks included in the Robitaille and Isella samples. \label{generalresults2} } 

\end{figure*}


\begin{table*}[th!]
\begin{minipage}{18.4cm}
\caption{Summary of the stellar and disk properties for the best-fitting dust models for our 13 disks. \label{srcprop} } 

\begin{minipage}{1.25cm}
\hspace{1.25cm}
\end{minipage}
\begin{minipage}{14cm}
\begin{tabular}{l|cccccc|ccccc} 
\hline 
\hline 
\noalign{\smallskip}

Source & \multicolumn{6}{c|}{Robitaille Model} & \multicolumn{5}{c}{Isella Model} \\
~ & $A_{\rm v}$ & M$_{\star}$ & M$_{\rm d}$ & R$_{\rm d}$ & $i$ & $F_{\rm 1.3mm}$ & M$_{\star}$ & M$_{\rm d}$ & R$_{\rm d}$ & $i$ & $F_{\rm 1.3mm}$  \\ 
~ & [mag] & [M$_{\odot}$] & [10$^{-2}$\,M$_{\odot}$] & [AU] & [$^{\circ}$] & [mJy] & [M$_{\odot}$] & [10$^{-2}$\,M$_{\odot}$] & [AU] & [$^{\circ}$] & [mJy] \\
\hline
CI Tau   & 1.80 (1.80) &  0.4 & 1.9 & 91  & 57 & 105 & -   & -	& -   & -  & -                \\
CW Tau   & - &  -   & -   & -   & -  & -   & -   & -	& -   & -  & -                \\
CY Tau   & 0.05 (0.03) &  0.9 & 1.4 & 97  & 32 & 23  & 0.4 & 6.92 & 197 & 51 & 117\,$\pm$\,20   \\
DG Tau   & 1.60 (1.41) &  1.5 & 3.9 & 158 & 50 & 254 & 0.3 & 41.7 & 89  & 18 & 317\,$\pm$\,28   \\
DN Tau   & 0.49 (0.49) &  0.6 & 1.0 & 92  & 32 & 39 & 0.4 & 1.86 & 125 & 30 & 93\,$\pm$\,8     \\
DO Tau   & 4.88 (2.23) &  0.8 & 3.2 & 104 & 57 & 103 & -   & -	& -   & -  & -                \\
DQ Tau   & -    &  -   & -   & -   & -  & -   & -   & -	& -   & -  & -                \\
DR Tau   & 0.51 (0.51) &  0.8 & 3.2 & 104 & 18 & 103 & 0.4 & 6.31 & 86  & 37 & 109\,$\pm$\,11   \\
FT Tau   & 0.00 (  -  )  &  0.2 & 0.8 & 62  & 50 & 36 & -   & -	& -   & -  & -                \\
GO Tau   & - &  -   & -   & -   & -  & -   & 0.6 & 7.10 & 670 & 25 & 57\,$\pm$\,8     \\
IQ Tau   & 1.44 (1.44) &  1.0 & 3.0 & 101 & 63 & 74 & -   & -	& -   & -  & -                \\
UZ Tau   & -    &  -   & -   & -   & -  & -   & 0.3 & 4.79 & 260 & 43 & 126\,$\pm$\,12   \\
V806 Tau & -    &  -   & -   & -   & -  & -   & -   & -	& -   & -  &   \\
\hline
\noalign{\smallskip}
\end{tabular}
\end{minipage}
\end{minipage}
\hfill

\begin{minipage}{18.4cm}
\begin{tiny}
\textbf{Notes.} For the Robitaille models, the visual interstellar extinction $A_{\rm v}$ magnitudes are for the best model fit, only using the measured values from \citet{white2001} as an input parameter to the SED fitting tool (in parenthesis). The M$_\star$ values listed for both models are the derived values for each method. For the Isella models, the outer dust radius $R_{\rm d}$ is from Table 5 in \citet{isella2009}, defined as where the disk becomes optically thin to the stellar radiation. For UZ Tau, the Isella model in fact concerns only the spectroscopic binary component UZ Tau E. The 1.3\,mm continuum fluxes ($F_{\rm 1.3mm}$) listed for our best-fitting Robitaille model are the SED values of the fit; and for the Isella models we list the resolved (interferometric) dust continuum measurement.
\end{tiny}
\end{minipage}
\end{table*}


Our second approach uses a model by \citet{isella2009}, which explicitely takes into account the spatial distribution of the millimeter continuum emission observed with the CARMA interferometer. These authors approximate the vertical temperature structure of the disk with the two-layer description by \citet{chiang1997}, and fit the grain size and opacity to resolved 1.3\,mm data. In our adaptation of their models, we omit the hot surface layer because it contains insignificant amounts of molecular gas, and we extract the disk's interior temperature from Fig.~7 of \citet{isella2009}. The surface density is obtained from their Eq.~9 and Table~5. We then calculate the local hydrodynamic equilibrium scale height following Eqs.~3 and 4 from \citet{hughes2008}. We truncate the models at the transition radius, defined by \citet{isella2009} as the location where the disk becomes optically thin to the stellar radiation (Sect.~\ref{dgtau} discusses the effect of extending the disk further).

Figs.~\ref{generalresults1} and \ref{generalresults2} show the resulting temperature and density structures for our sources. For some sources, only one type of model is shown because of the availability of models in \citet{robitaille2007} and \citet{isella2009}. Table \ref{srcprop} summarizes the parameter fits for both models. As can be immediately seen for the four sources for which both a Robitaille and an Isella model are available (i.e.~CY Tau, DG Tau, DN Tau, and DR Tau), widely different disk descriptions apparently provide equally good fits to the continuum data. 

The vertical height of the Robitaille models is often 4--6 times \emph{smaller} than the Isella models. The radial extents of the disks are comparable for both approaches to within a factor of $<$\,2. The resulting masses differ by factors of 2--10, with the Isella models always producing more massive disks. The temperature profiles of the Robtaille models are more detailed (by definition) while the Isella models at large radii are close to isothermal. In general, the temperature between the two models differs by a factor of 3--5 in the midplane. In addition to these parameters, other factors that will affect the emergent lines are the disk inclinations and stellar masses, both of which influence the widths of the lines.

To calculate the resulting molecular line emission, we populate each model disk with gas using the standard 100:1 gas-to-dust ratio, a mean molecular weight of 4.008\,$\times$\,10$^{-24}$\,g, and constant H$_2$ relative abundances of 1\,$\times$\,10$^{-9}$ for HCO$^+$, 2\,$\times$\,10$^{-11}$ for HCN, and 1\,$\times$\,10$^{-9}$ for CN, based on the fractional values in the warm molecular layer for the theoretical models of \citet{aikawa2002} and \citet{zadelhoff2003}. In regions where the dust temperature (assumed equal to the gas temperature) drops below the CO ice evaporation temperature of 20\,K, these abundances are reduced by a factor 10$^3$ to account for CO freeze-out. When the H$_2$ number density drops below 10$^{3}$\,cm$^{-3}$, the molecular abundances are set to zero, effectively defining the disk edge. 

The gas kinematics follow a cylindrical Keplerian velocity field with stellar masses corresponding to each model fit, as indicated in Table \ref{srcprop}. A Doppler broadening factor of 0.15\,km\,s$^{-1}$ is also factored into the line calculations. The statistical equilibrium molecular excitation and line radiative transfer was solved using the RATRAN code \citep{hogerheijde2000b}, and the emerging line emission was convolved with the appropriate Gaussian beams. The resulting line profiles are plotted alongside the disk models in Figs.~\ref{generalresults1} and \ref{generalresults2}.

\subsection{Comparison of fixed-abundance models}
\label{general} 

As is immediately obvious from Figs.~\ref{generalresults1} and \ref{generalresults2}, the models have difficulty reproducing the detected emission lines. In four cases (CI Tau, DG Tau, DN Tau, and DO Tau) the models predict lines that are weaker than observed; and in two cases (GO Tau and UZ Tau), the models predict lines that are stronger than the detected line (for GO Tau) or the obtained upper limits (for UZ Tau). In the remaining four cases (CY Tau, DR Tau, FT Tau, and IQ Tau), the predicted lines are consistent with the achieved upper limits, although for CY Tau the Isella model produces an HCO$^+$ line that violates the upper limit. 

Where both a Robitaille model and an Isella model are available, the Isella model always produces lines that are
stronger and narrower than those of the Robitaille models. The larger stellar masses, by factors 2--5, of the latter, and the larger disk masses of the former, by factors 2--10, contribute to this difference. Inclination also plays a significant role, with more face-on orientations leading to stronger lines (cf.~DO Tau and DR Tau, which are fit with the same Robitaille disk structure, but have respective inclinations of $i=57^\circ$ and $i=18^\circ$; DR Tau has predicted lines stronger by a factor of $\sim$3). Interestingly, of the six sources from our sample modeled by \citet{isella2009}, the three with \emph{detected} line emission have $i\le 30^\circ$ (DG Tau, DN Tau, and GO Tau), while the sources with $i>30^\circ$ (CY Tau, DR Tau, and UZ Tau) are undetected (all have similar M$_\star$). Perhaps the narrower line profiles have helped to make these sources detectable. 

Given the general mismatch between the predicted line intensities and widths, and the observations, it is not possible to draw conclusions about the HCO$^+$, HCN, or CN abundances this way. Simply scaling up or down the abundance will not result in a match (to the line shape); only for GO Tau do the abundances appear to lie within a factor of a few above the true values. Furthermore, as illustrated by the case of DR Tau, two \emph{different} disk models (but with more comparable M$_{\rm d}$, $R_{\rm d}$, and $F_{\rm 1.3mm}$ values) produce very similar lines. These degeneracies make it difficult to derive reliable conclusions about the HCO$^+$, HCN, and CN abundances. Instead, more detailed modeling of individual sources may be required.



\begin{figure*}
\centering
\includegraphics[width=9cm,angle=-90]{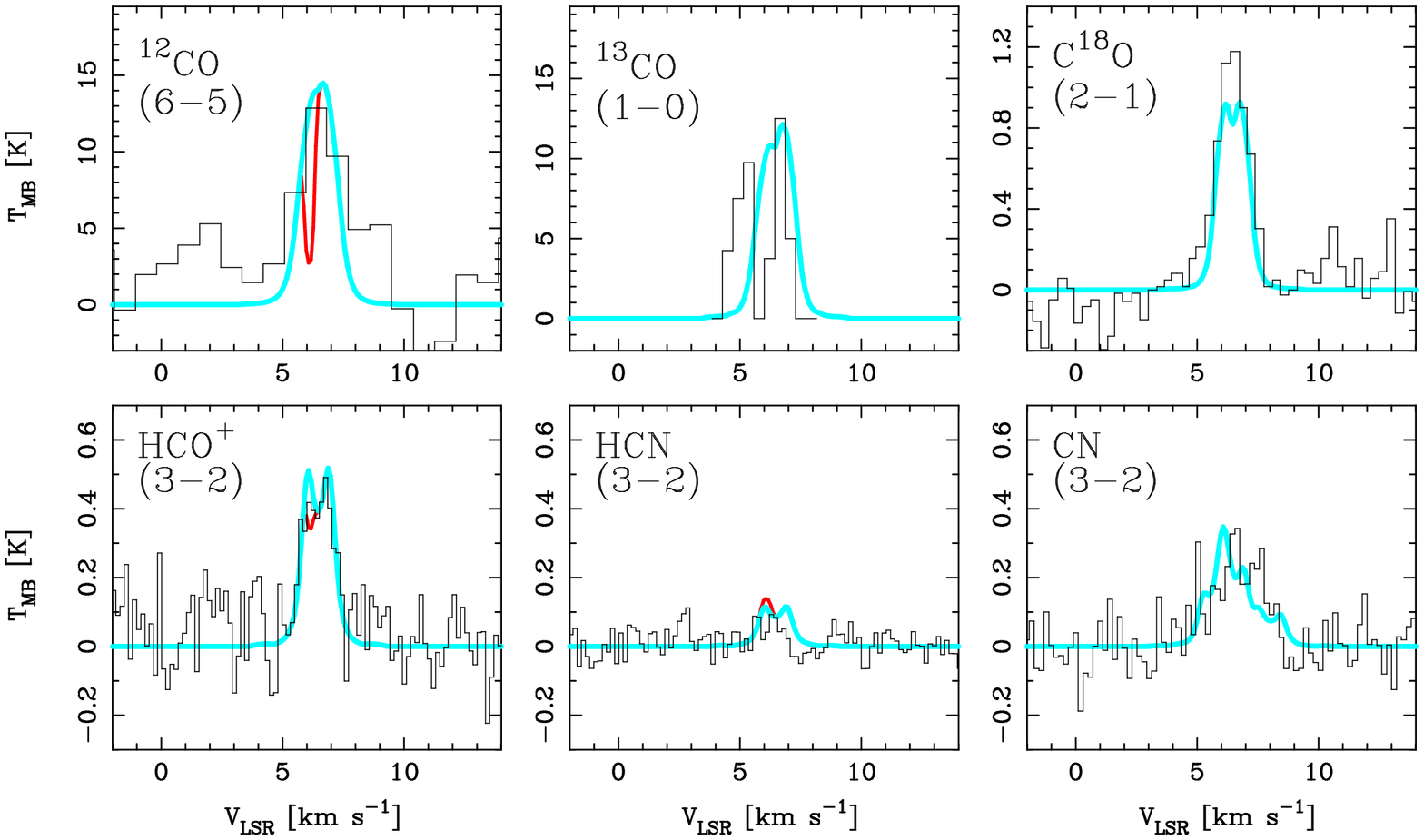}
\caption{Line predictions for our best-fitting model of DG Tau. The $^{12}$CO\,(6--5) and C$^{18}$O\,(2--1) data are taken from \citet{schuster1993} where their $T_{R}^{*}$ scale is equivalent to our $T_{\rm mb}$ scale. The $^{13}$CO\,(1--0) is from \citet{kitamura1996a}, and since we were unable to establish a conversion from Jy to K, we compare the \textit{line shape} only. We have indicated with a thick blue line our fits for a 600\,AU disk at an inclination of 25$^{\circ}$. Indicated in red is how the ($^{12}$CO, HCO$^+$, and HCN) line profiles are affected by the absorption or excess emission from a cold, foreground cloud at a radial velocity of 6.1\,km\,s$^{-1}$. No differences are visible for the $^{13}$CO or C$^{18}$O lines. The effect of the foreground cloud on the CN line is not available; but the expected hyperfine line-splitting is calculated. \label{incl25} }
\end{figure*}


\subsection{The specific case of DG Tau} 
\label{dgtau} 

Of all the sources in our sample, DG Tau offers the best case to obtain a detailed model. Its 1.3\,mm continuum is brighter by a factor $\geq$\,2.5 than any of the other sources and it emits the strongest HCO$^+$ and CN lines. An extensive literature also exists on DG Tau describing sub-millimeter single-dish \citep{schuster1993,mitchell1994} and interferometer data \citep{kitamura1996a,kitamura1996b,dutrey1996,testi2002,isella2009}. Its millimeter continuum emission is compact with $\sim$80\% originating from within 95\,AU \citep{isella2009}. The best-fit Robitaille model obtained in the previous section (see Table \ref{srcprop}, Robitaille database model number 3017659) matches other literature estimates of the DG Tau disk parameters (Table \ref{dgtaulit}). Only the disk mass (M$_{\rm d}$\,=\,0.042\,M$_\odot$) for the Robitaille model is significantly larger than the literature values (0.015--0.025\,M$_\odot$). In contrast, the Isella model offers a disk mass value 10$\times$ greater (0.4\,M$_\odot$). The Isella model also overestimates the stellar mass (M$_\star$\,=\,1.5\,M$_\odot$) with respect to most literature values where M$_\star$ ranges from 0.3 to 0.8\,M$_\odot$; the Robitaille model yields 0.3\,M$_\odot$ in comparison. To model DG Tau we settle on a central stellar mass of 0.8\,M$_{\odot}$, which is on the high end of the literature values for this object but provides the best fit to the line profiles, as discussed below.

Figure \ref{incl25} reproduces the $^{12}$CO\,(6--5), $^{13}$CO\,(1--0), and C$^{18}$O\,(2--1) line observations of \citet{schuster1993} and \citet{kitamura1996a}. The bright CO isotopologues suggest a significant gas reservoir. Interferometric imaging of the $^{13}$CO\,(1--0) and (2--1) lines in the literature reveal a gaseous disk structure of 600\,AU in extent \citep{kitamura1996a,kitamura1996b,testi2002}, about 4--6 times the size inferred from dust emission. Accordingly, we adopt 600\,AU for the outer radius of the gas disk, extrapolating the initial Robitaille model outwards. In addition, the $^{13}$CO\,(2--1) emission observed by \citet{testi2002} is consistent with Keplerian rotation around a star of 0.67\,$\pm$\,0.25\,M$_{\odot}$, oriented perpendicular (within $\pm$15$^{\circ}$) to the highly collimated jet system, which is inclined 38$^{\circ}$ with respect to the line of sight \citep{eisloffel1998}. Therefore, we adopt an inclination of 25$^\circ$, which also gives the best fit to the C$^{18}$O\,(2--1) line profile. We already show in Fig.~\ref{incl25}---and we will discuss later in this section---the minimal effect from intervening cloud or remnant envelope material in this line.

With the stellar mass (M$_\star$\,=\,0.8\,M$_\odot$), inclination ($i$\,=\,25$^\circ$), and outer gas radius ($R_{\rm d}$\,=\,600\,AU) now fixed, we use a constant fractional CO abundance of 2\,$\times$\,10$^{-4}$ and C$^{18}$O abundance of 4\,$\times$\,10$^{-7}$ (except when $T$\,$<$\,20\,K, where an abundance 10$^3$ times lower is used) to calculate the simulated line profiles with RATRAN using the extended and modified Robitaille model. To fit the lines, particularly C$^{18}$O, we find that we need to increase the gas temperature by a factor 1.7, suggesting that the line emission originates in layers where $T_{\rm gas}$\,$>$\,$T_{\rm dust}$. We adopt this same scaling for the gas temperature for all species, but neglect its effect on the scale height.

For these model parameters, the HCO$^+$ line can be reproduced very well for a disk-averaged abundance of 2\,$\times$\,10$^{-11}$ with respect to H$_{2}$. This is lower than theoretical predictions for the warm emission layer in many T~Tauri disks \citep{aikawa2002,zadelhoff2003}, but not unlike the beam-averaged fractional abundances ($\sim$10$^{-11}$--10$^{-12}$) inferred from observations of disks around several high-mass (Herbig Ae/Be) stars \citep[see][]{thi2004}. A low mean abundance is also especially surprising given DG Tau's powerful jets, which emit significant X-ray radiation \citep{gudel2008b}. Apparently, the disk is sufficiently shielded to retain a low ionization degree. An upper limit for the HCN abundance of $5\times 10^{-12}$ is found, while a value of $8\times 10^{-10}$ is obtained for CN. This high CN/HCN ratio of $>$\,160 suggests efficient HCN dissociation in the bulk of the disk, which is more consistent with its mid-infrared characterization indicating fewer small grains and some dust settling. Here we note the importance of using \emph{abundance} ratios rather than line \emph{intensity} ratios (which contain opacity and excitation effects): the ratio of the integrated intensities of CN/HCN used in Sect.~\ref{results} is $>$\,12, a factor of 13 smaller than the underlying abundance ratio found here, albeit consistent.


\begin{table}[b!]
\begin{minipage}[]{\linewidth}
\caption{DG Tau star and disk properties from the literature to compare with the two disk models. \label{dgtaulit} }
\centering
\begin{tabular}{lcccl}
\hline
\hline
\noalign{\smallskip}
Property & Isella & Robitaille & Literature & References \\
& Model & Model & Range &  \\
\hline
\noalign{\smallskip}
Sp.~Type & M0 & - & K1--K7 & 2,~10,~12,~23\\
M$_\star$~[M$_{\odot}$] & 0.3 & 1.48 & 0.3--0.8 & 4,~5,~17,~20\\
Age~[Myr] & 0.1 & 1.89 & 0.3--2.0 & 5,~11,~13  \\
T$_{\rm eff}$~[K] & 3890 & 4560 & 3890--4395 & 5,~7,~11,~17 \\
L$_\star$~[L$_{\odot}$] & 1.70 & - & 1.07--3.62 & 1,~4,~5,~18 \\
R$_\star$~[R$_{\odot}$] & 2.87 & 2.427 & 2.13--2.8 & 3,~11,~17 \\
$\dot{\rm M}$~[M$_{\odot}$\,yr$^{-1}$] & 4.1e$^{-7}$ & 4.5e$^{-7}$ & 1.2--20e$^{-7}$ &
11,~17,~18  \\
M$_{\rm d}$ [10$^{-2}$\,M$_{\odot}$] & 41.7 & 3.9 & 1.5--2.51 & 2,~6,~9,~13 \\
R$_{\rm dust}$ [AU] & 89 & 158 & 80--300 & 9,~16,~19 \\
$R_{\rm gas}$ [AU] & 160 & 600 & 600 & 16,~24 \\
$i$~[$^{\circ}$] & 18 & 25 & 18-90 & 3,~9,~14,~16 \\
A$_{\rm v}$ [mag] & - & 1.6 & 1.41--1.6 & 10, 22 \\
Dist.~[pc] & 140 & 140 & 128--156 & 10, 21 \\
\noalign{\smallskip}
\hline
\noalign{\smallskip}
\end{tabular}
\end{minipage}

\hfill

\begin{minipage}[h]{\linewidth}
\begin{tiny}
\textbf{References.} (1)~\citet{akeson2005}; (2)~\citet{andrews2005}; (3)~\citet{appenzeller2005}; (4)~\citet{basri1991}; (5)~\citet{beckwith1990};
(6)~\citet{beckwith1991}; (7)~\citet{bouvier1995}; (8)~\citet{cieza2005}; (9)~\citet{dutrey1996}; (10)~\citet{furlan2006}; (11)~\citet{hartigan1995}; (12)~\citet{hessman1997}; (13)~\citet{honda2006}; (14)~\citet{isella2009};
(15)~\citet{kenyon1995}; (16)~\citet{kitamura1996b}; (17)~\citet{mohanty2005}; (18)~\citet{muzerolle2003}; (19)~\citet{rodmann2006}; (20)~\citet{tamura1999};
(21)~\citet{vinkovic2007}; (22)~\citet{white2001}; (23)~\citet{white2004}; and (24)~\citet{testi2003}.
\end{tiny}
\end{minipage}

\end{table}

\begin{table}[t!]
\begin{minipage}[]{\linewidth}
\caption{Determined abundances for our best-fit DG Tau model. \label{abundances} }
\centering
\begin{tabular}{lcc}
\hline
\hline
\noalign{\smallskip}
Molecule & DG Tau & Theoretical \\
\hline
\multicolumn{3}{c}{\textit{Disk Fractional Abundances (w.r.t. H$_{2}$)\,$^{a}$}} \\
\hline
\noalign{\smallskip}

$^{12}$CO & 2.0\,$\times$\,10$^{-4}$  & 1.0\,$\times$\,10$^{-4}$\\
$^{13}$CO & 3.3\,$\times$\,10$^{-6}$  & 1.7\,$\times$\,10$^{-6}$ \\
C$^{18}$O & 4.0\,$\times$\,10$^{-7}$  & 2.0\,$\times$\,10$^{-7}$ \\
HCO$^{+}$ & 2.0\,$\times$\,10$^{-11}$ & 1--100\,$\times$\,10$^{-11}$ \\
HCN       & 5.0\,$\times$\,10$^{-12}$ & 1--100\,$\times$\,10$^{-11}$\\
CN        & 8.0\,$\times$\,10$^{-10}$ & 1--100\,$\times$\,10$^{-11}$ \\

\hline
\multicolumn{3}{c}{\textit{Cloud Fractional Abundances (w.r.t. H$_{2}$)\,$^{b}$}} \\
\hline
\noalign{\smallskip}

$^{12}$CO & 2.0\,$\times$\,10$^{-4}$ & 8.0\,$\times$\,10$^{-5}$ \\
$^{13}$CO & 3.3\,$\times$\,10$^{-6}$ & - \\
C$^{18}$O & 4.0\,$\times$\,10$^{-7}$ & - \\
HCO$^{+}$ & 8.0\,$\times$\,10$^{-9}$ & 8.0\,$\times$\,10$^{-9}$ \\
HCN       & 8.0\,$\times$\,10$^{-9}$ & 4--20\,$\times$\,10$^{-9}$ \\
CN\,$^{c}$        & -                & 3--30\,$\times$\,10$^{-9}$ \\

\hline
\multicolumn{3}{c}{\textit{Cloud Column Densities (cm$^{-2}$)}} \\
\hline
\noalign{\smallskip}

$^{12}$CO & 6.0\,$\times$\,10$^{16}$ & $\cdots$ \\
$^{13}$CO & 1.0\,$\times$\,10$^{15}$ & $\cdots$ \\
C$^{18}$O & 1.2\,$\times$\,10$^{14}$ & $\cdots$ \\
HCO$^{+}$ & 2.4\,$\times$\,10$^{12}$ & $\cdots$ \\
HCN       & 2.4\,$\times$\,10$^{12}$ & $\cdots$ \\
CN        & -                        & $\cdots$ \\

\hline
\noalign{\smallskip}
\end{tabular}
\end{minipage}
\hfill
\begin{minipage}[h]{\linewidth}
\begin{tiny}
\textbf{Notes.} (a)~The DG Tau abundances are constant throughout the disk (or disk-averaged), whereas the theoretical disk abundances from \citet{aikawa2002} and \citet{zadelhoff2003} represent the ranges expected in the warm molecular layers only; (b)~The theoretical cloud values are from \citet{terzieva1998}; (c)~The online RADEX program does not yet include CN (to calculate the cloud contributions).
\end{tiny}
\end{minipage}

\end{table}

\begin{figure*}[tbh!] 
\centering 
\includegraphics[width=16cm]{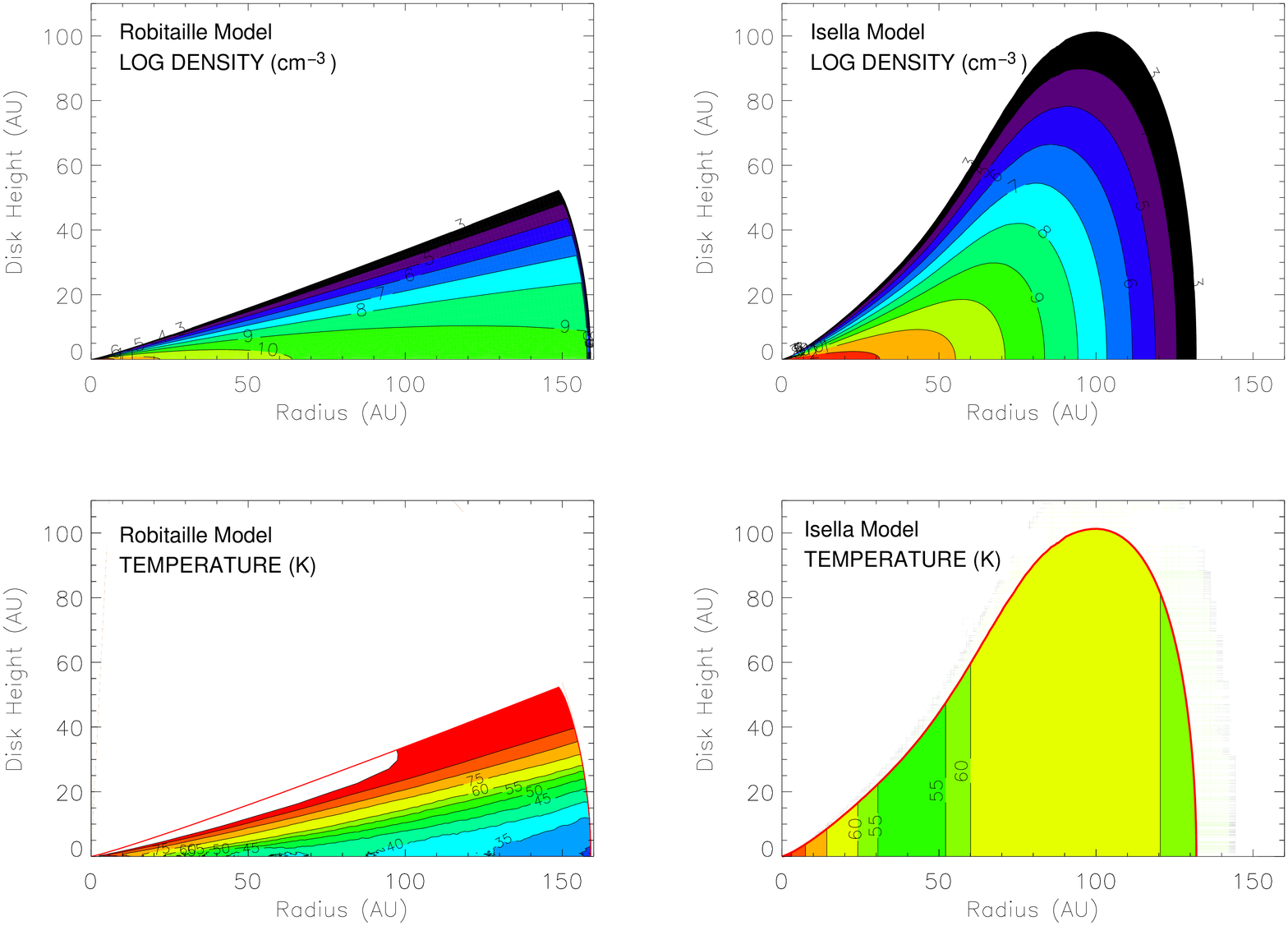} 
\caption{The radial and vertical structures for the \emph{inner} 160\,AU of each disk model, for direct comparison. The panels at left show the Robitaille model (without the extended gas reservoir), and the panels at right show the full Isella model (with exponential taper). The upper panels indicate the temperature structures, and the lower panels compare the density structures for each model. (Color plots are available online) \label{diskstructures} } 
\end{figure*}

\begin{figure*}[tbh!] 
\centering 
\includegraphics[width=9cm,angle=-90]{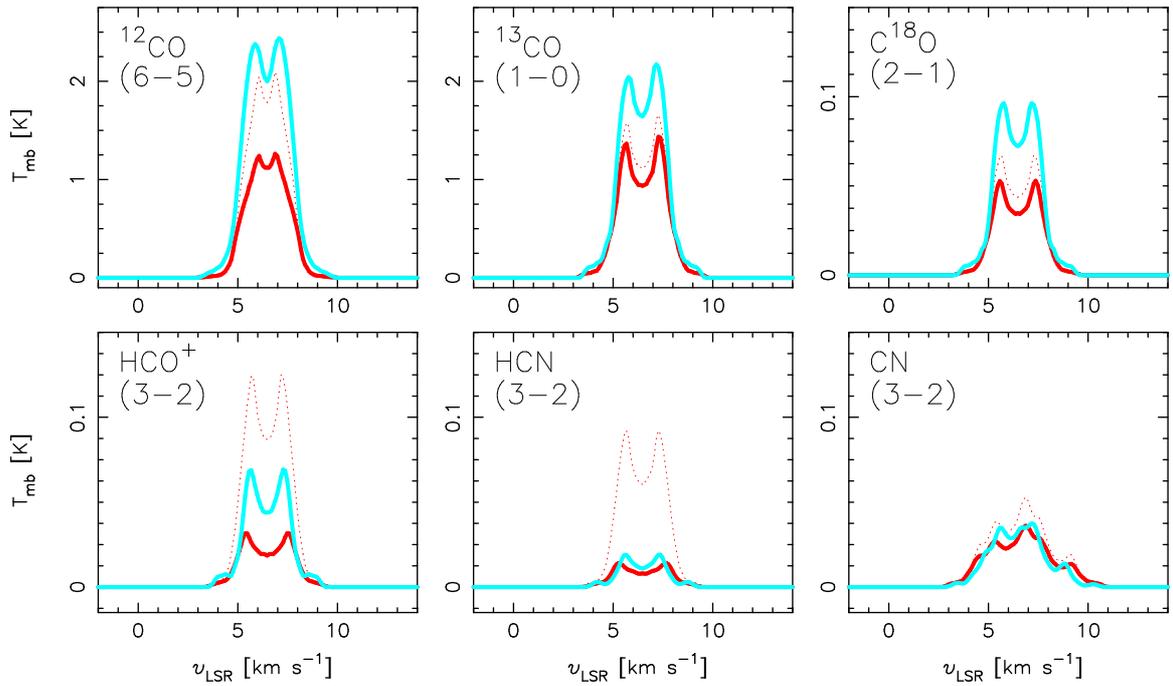} 
\caption{Similar to Figs.~\ref{generalresults1} and \ref{generalresults2}, the line predictions for the two models for the \emph{inner} 160\,AU, assuming identical abundances, $T_{\rm kin}$, M$_\star$, and $i$, in order to uniformly compare the predicted emission. A solid blue line represents the Robitaille solution, whilst a red line is the Isella prediction. The corresponding temperature and density structures are given in Fig.~\ref{diskstructures}. In addition, as a dashed red line, we plot the most extreme solution to the Isella model, based on the provided error bars and the largest possible fractional abundances predicted for disks. For both models, the inner 160\,AU structure emits only a small fraction of the total emission necessary to recreate the observed lines, but the two predictions are very similar. \label{comparison} } 
\end{figure*}

The disk abundances above have been derived using the density and temperature structure of the best-fitting Robitaille model, with noted modifications. The results are summarized in Table \ref{abundances}. Substituting the description of Isella instead (but keeping the same values for M$_\star$ and $i$), the emergent line intensities are lower by up to a factor of $\sim$16 for HCO$^+$ for the same abundances, and by smaller factors of $\sim$5 and $\sim$10 for HCN and CN, respectively. The large line intensity differences result also from the differences in $R_{\rm gas}$ for each model. Whereas the Robitaille power-law model for the temperature and density description (and sharp outer dust edge) leads easily to extrapolation to larger radii, the Isella model, with its exponentially tapered density structure, does not. Thus, to compare the two disk structures in a uniform way, we plot the temperature and density profiles for each model for the \emph{inner} 160\,AU \emph{only} in Fig.~\ref{diskstructures}. Then, in Fig.~\ref{comparison} we plot the predicted lines. Unlike Figs.~\ref{generalresults1} and \ref{generalresults2}, we now assume identical gas kinematic properties (M$_\star$, $i$, and $R_{\rm gas}$). The predicted lines for each model in Fig.~\ref{comparison} are now strikingly similar. The differences in the two emerging line intensities are lower by up to a factor of 2 for HCO$^+$ and much closer for HCN and CN. This suggests that the large CN/HCN ratio of $>$\,160 found for the underlying abundances is independent of the temperature and density details of the adopted model (for the inner 160\,AU); and leaves only the missing details of the input radiation field uninvestigated.

Finally, DG Tau is not an isolated source. The environment around the star is dominated by an optical jet \citep{eisloffel1998}, a strong molecular outflow \citep{mitchell1994}, an expanding circumstellar envelope \citep{kitamura1996b}, and intervening cloud material (this work). The result of this confused environment is most clearly evident when comparing the $^{12}$CO\,(3--2) line observations presented in \citet{schuster1993} and \citet{mitchell1994}, which exhibit equally bright line intensities and significant wings at the on-source position and three separate offset positions. We chose to omit the $^{12}$CO observations from the fits in Fig.~\ref{incl25}, since they distract from the disk emission. However, we do determine that the $^{12}$CO lines are about 3$\times$ stronger than predicted for our disk model, suggesting contributions from a surrounding cloud with a CO column density of $N_{\rm CO}$\,$\approx$\,6\,$\times$\,10$^{16}$\,cm$^{-2}$ and line width of 0.3\,km\,s$^{-1}$ for an adopted cloud temperature of 25\,K and H$_2$ number density of $10^4$\,cm$^{-3}$ (or $N_{\rm CO}$\,$\approx$\,1\,$\times$\,10$^{16}$\,cm$^{-2}$ for an H$_2$ number density of 10$^5$\,cm$^{-3}$). For these typical cloud densities, we do not expect significant HCO$^+$\,(3--2), HCN\,(3--2), or CN\,(2--1) emission (or absorption); and both Figs.~\ref{hcodetections} and \ref{incl25} confirm this.


To determine those line contributions from the cloud, plotted in Fig.~\ref{incl25}, as well as the cloud fractional abundances and column densities listed in Table \ref{abundances}, we used the RADEX\footnote{RADEX is an online, one-dimensional, non-LTE radiative transfer code developed by \citet{vandertak2007} to calculate the molecular line intensity, opacity, and excitation temperature. For more information, see http://www.sron.rug.nl/$\sim$vdtak/radex/radex.php} online calculator. We modeled the cloud as a cold intervening layer moving with a radial velocity of 6.1\,km\,s$^{-1}$. We note that in many of the literature observations, strong absorption is seen near a velocity of 5.8--6.2\,km\,s$^{-1}$, with several studies reporting these values as the source velocity. We emphasize here that our observations of HCO$^+$, with a critical density 3 orders of magnitudes larger than the CO observations in the literature, are a much  better tracer of the disk content, establishing the source radial velocity at 6.47\,km\,s$^{-1}$. In addition, we confirm that the emerging HCO$^{+}$ line predicted for our intervening cloud fits the detections observed at both DG Tau offset positions ($\sim$0.25\,K peak centered at 6.1\,km\,s$^{-1}$, refer back to Fig.~\ref{hcodetections}). The effect of the intervening material on the observed $^{12}$CO, HCO$^+$, and HCN lines is shown in Fig.~\ref{incl25}, while $^{13}$CO and C$^{18}$O exhibit no differences, and the CN cloud predictions are not yet available in the RADEX program.

\subsection{Notes on individual sources} 
\label{others} 

\textbf{V806 Tau}, also called Haro 6--13, has the second strongest HCO$^{+}$ line after DG Tau, but is undetected in HCN and CN. It is a single M0 star \citep{rebull2010,luhman2010}. With an $A_{\rm v}$\,=\,11.2 \citep{furlan2006}, its optical extinction is much larger than our other sources, which might help explain its unique silicate emission feature and positive spectral slope. \citet{furlan2006} comment that the silicate features are reminsicent of transitional disks, although it also possesses a high mass accretion rate \citep{white2004}. Some extended HCO$^+$ emission is apparent in Fig.~\ref{hcodetections}. However, our single-dish radial velocity of 5.40\,km\,s$^{-1}$ is consistent with the value of 5.10\,km\,s$^{-1}$ from CO interferometric observations \citep{schaefer2009}. V806 Tau's low disk mass of 0.01~M$_\odot$ \citep{andrews2005,honda2006} and large 400\,AU radius \citep{robitaille2006} suggest that much of the disk of V806 Tau should be UV illuminated, in contrast to what our observed upper limit for CN would suggest. 

\textbf{GO Tau} shows complex line profiles, with peaks at a $\upsilon_{\rm LSR}$ of 4.4, 5.1, 5.5, and 6.3 km s$^{-1}$. \citet{thi2001} report $^{12}$CO emission peaks at 5.2, 5.5, 6.2, and 7.1 km s$^{-1}$, and $^{13}$CO emission at 4.3 and 7.0 km s$^{-1}$. They attribute the 5.5 and 6.2 km~s$^{-1}$ components to surrounding cloud emission. However, we assign the emission between 4--6\,km\,s$^{-1}$ to the disk of GO Tau, as suggested by the interferometric observations of \citet{schaefer2009}
and \citet{andrews2007phd}. For GO Tau, the accretion rate is very low \citep{hartmann1998}, and the amount of dust settling is small if we draw upon its mid-infrared spectral slope, suggesting lower rates of UV photodissociation and ionization should be occurring. However, it exhibits one of the brightest HCO$^+$ lines, in stark contrast to the rest of the sample since its fainter 1.3\,mm continuum flux just straddles our cutoff and yet the source also appears to be associated with a large, and dense, gas reservoir.

\textbf{DR Tau} also has a complicated circumstellar environment, with $^{12}$CO emission lines at 6.8, 9.1, 10.0, and 10.3 km s$^{-1}$ and $^{13}$CO emission at 6.9 km s$^{-1}$ and near 11\,km\,s$^{-1}$ \citep{thi2001}. SMA interferometric observations by \citet{andrews2007phd} also show strong emission centered on 10.5\,km\,s$^{-1}$, suggesting that this is the correct source velocity and lending credence to our assertion that the CN emission likely originates in the disk. Our CN spectrum shows a triple-peaked 5.4$\sigma$ feature centered at 11.2 km~s$^{-1}$ (Fig.~\ref{cnfig}). Our model results suggest that `standard' HCO$^+$ and HCN abundances are consistent with the non-detections of the lines. The CN detection in the absence of the other lines, on the other hand, indicates a significant CN enhancement.

\textbf{CW Tau} is clearly surrounded by dense cloud material, as witnessed by the equally strong HCO$^+$ emission on- and off-source. Our HCO$^+$ observations (see Fig.~\ref{hcodetections}) illustrate how, in a crowded star-forming region, measurements at offset positions can be both relevant and useful even for molecular species that preferentially trace much denser material.

\textbf{CY Tau, DQ Tau, IQ Tau,} and \textbf{UZ Tau} do not show gas emission lines in our data. Interestingly, they span the full dust classification and morphological sequence of \citet{furlan2006} with CY Tau and DQ Tau showing rather flat, decreasing mid-infrared SEDs, and IQ Tau and UZ Tau showing evidence for a small grain population. The mass accretion rates -- which contribute to the stellar UV excess -- range from very low (10$^{-9}$\,M$_\odot$\,yr$^{-1}$) for CY Tau to average (10$^{-7}$\,M$_\odot$\,yr$^{-1}$) for DQ Tau and UZ Tau \citep{gudel2007a}. However, both DQ Tau and UZ Tau E are spectroscopic binaries that exhibit pulsed accretion events on periods of weeks \citep{basri1997}, and their higher reported accretion rates may overestimate the average, quiescent values.

\section{Discussion}
\label{discussion}

\begin{figure}[t!]
\centering
\includegraphics[width=8cm]{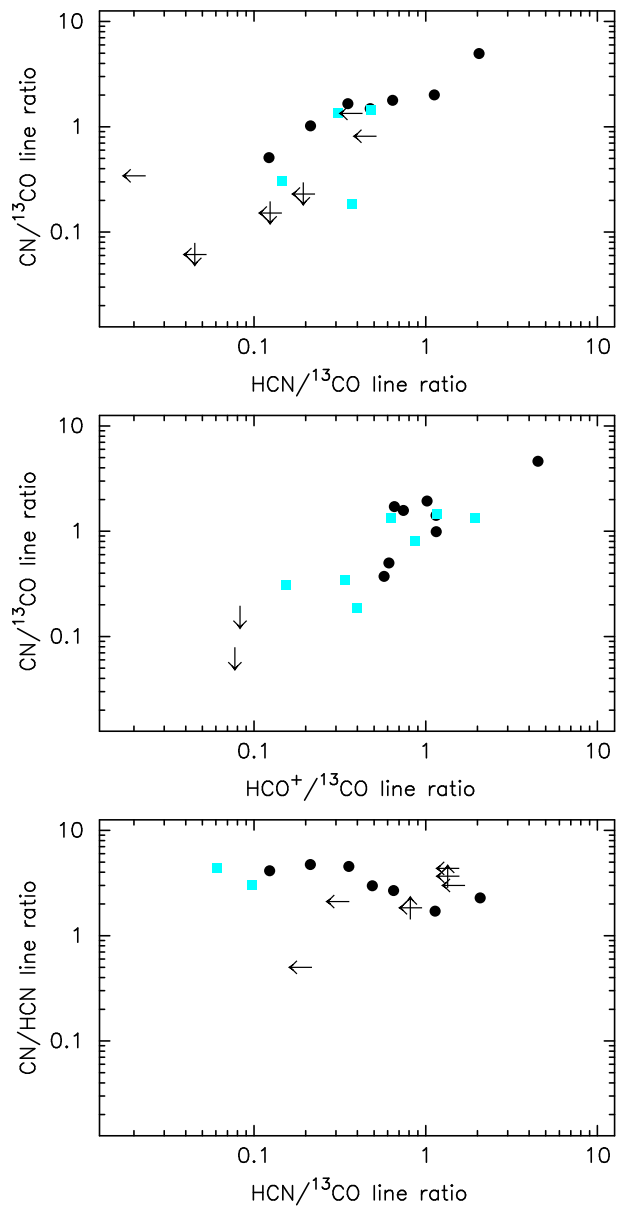}
\caption{Integrated intensity line ratios for HCO$^{+}$, HCN, CN, and $^{13}$CO plotted alongside other disk systems around PMS stars in the literature. Circles represent data points extracted from Fig.~3 of \citep{kastner2008b}, squares represent data points from our complete Taurus sample, and black arrows represent upper/lower limits. \label{lineratios} } 
\end{figure}

To return to previous work, \citet{kastner2008b} showed plots of HCO$^+$, HCN, and CN line ratios for several PMS stars, showing a tentative correlation between the HCO$^+$, HCN, CN, and $^{13}$CO line ratios. In Fig.~\ref{lineratios} we reproduce their plots and their data points (without error bars), and add our own line ratios. Overall, we find that the trends persist: CN is typically stronger than HCN, CN is also stronger than HCO$^+$, and the photodissociation rate (as probed by CN/HCN) is roughly constant regardless of the HCN relative line strength. Our contributed line ratios consist largely of upper and lower limits and therefore do not specifically challenge or confirm the trends by probing different regions in the plots. While this complete PMS sample in Fig.~\ref{lineratios} includes sources covering a range in age, mass, and radiation field, the line strengths of HCO$^+$, HCN, and CN relative to one another do reveal the importance of the ongoing UV photodissociation and X-ray ionization processes in these disks. The results, however, are still limited by small numbers statistics, numerous upper limits, and different rotational transitions (that may trace different regions of the disk). 

The motivation of this study was to determine whether disks with a higher degree of dust settling, or with a decreased dust content, have higher abundances of CN and HCO$^+$ reflecting larger degrees of photodissociation and photoionization. Our data are inconclusive. In Sect.~\ref{trends} we found that the HCO$^+$, HCN, and CN line fluxes (or their ratios) do not depend on any other disk or stellar parameter such as millimeter flux, infrared slope, silicate feature strength, stellar spectral type, etc. Sect.~\ref{modeling} shows that detailed SED-based models have intrinsic degeneracies that preclude straight-forward modeling. And even a detailed model, tailored to the case of DG Tau, does not provide unambiguous estimates of the HCO$^+$, HCN, and CN abundances.

This suggests two possible ways forward. In the first, spatially resolved observations of both the dust continuum and the line emission can be used to obtain \emph{in situ} measurements of the molecular abundances. ALMA will be a powerful instrument for an analysis like this. By addressing localized disk regions, rather than the emission integrated over the entire face of the disk, more simplified modeling approaches can be used (not unlike what is presently state-of-the-art analyses for photon dominated regions, PDRs). In addition, spatially resolved observations provide many more constraints on the underlying disk structure, such as the extent and surface density. This approach addresses the question of how molecular line emission and underlying disk structure are interrelated.


The second approach does not focus on the details of the underlying disk structure, but rather the details of the star's radiation field, which irradiates the disk. Accurate determination of the spectral type and luminosity, the UV and X-ray emission characteristics, and their time dependences should lead to a better understanding of the \emph{response} of the molecular gas reservoir to the incident radiation. High-resolution, high signal-to-noise observations are time-consuming but necessary for better stellar characterization of PMS stars, while veiling and other circumstellar environmental effects provide challenges that have been overcome \citep{merin2004,herczeg2008}. Recent studies have even argued that the continuum UV spectrum for T~Tauri stars must be much weaker, due to the total fraction of the stellar FUV flux that is emitted in the Ly$\alpha$ line alone. Since this fraction can range from 30\% up to 85\% (for TW Hya), Ly$\alpha$ provides an additional source of photodissociating power that varies from source-to-source and is extremely difficult to measure \citep{bergin2003}. Which of these two approaches is most fruitful will, of course, depend on which factor dominates the molecular line emission: the underlying disk structure or the stellar irradiation. Since these may be interrelated, both approaches may prove necessary.

\section{Summary}
\label{summary}

We surveyed 13 classical T~Tauri stars in low-$J$ transitions of HCO$^+$, HCN, and CN to compare the gas structure and chemical abundances within planet-forming disks that possess similar dust masses. We found, conversely, a wide variety of molecular gas properties. For this sample in the Taurus star-forming region, we report 6 new detections of HCO$^+$\,(3--2), 0 new detections of HCN\,(3--2), and 4 new detections of CN\,($J$\,=\,2--1). These data double the pool of previously known detections, bringing the total detection statistics toward the 21 brightest (at 1.3\,mm) disks in Taurus to: 14 for HCO$^+$, 5 for HCN, and 8 for CN. 

Overall the HCO$^+$, HCN, and CN line ratios for our Taurus disk sample are consistent with the trends identified toward other disks around PMS stars found in the literature, as initially plotted by \citet{kastner2008a,kastner2008b}. In general, CN is more prevalent than HCN, which suggests that the bulk of the detected emission originates in a UV photodissociating region. Additionally, the fractional molecular ionization ratio, as traced in only slightly denser regions by the HCO$^{+}$ line, is also enhanced. Both trends agree with the narrow emission lines observed in our sample, which trace the outermost regions where the disk is optically thin to \emph{both} the stellar and interstellar radiation fields. 

Despite this disk-to-disk agreement in the line ratios toward the general population of unresolved disks, the gas-line properties reveal \emph{no} observed chemical photoprocessing effects due to the dust properties or several stellar parameters, which was the main motivation for this research. We do not see CN and HCO$^{+}$ enhancements (via brighter lines) in sources whose mid-infrared spectral slope and silicate emission features indicate more grain growth and dust settling (leading to a lack of UV shielding). In addition, stellar parameters like X-ray luminosity do not seem to influence the observed line intensities of ionization tracers such as HCO$^+$.

The next step was to derive the underlying molecular abundances using two dust models in the literature that are then populated with theoretical values for the fractional molecular abundances. Models for the sample as a whole illustrate the importance of the M$_\star$\,sin\,$i$ and $R_{\rm gas}$ factors in gas-line abundance studies; parameters that are poorly constrained by the dust properties, but critical to proper line fits. Along these lines, we found during detailed modeling of the source DG Tau that the underlying abundances were less dependent on the temperature and density details in the adopted dust models. We conclude that better characterization of the stellar parameters (M$_\star$), the radiation field itself (UV and Ly$\alpha$), and spatially-resolved line observations ($R_{\rm gas}$ and $i$) are necessary to constrain the molecular gas content and evolution.  


Large sample statistics are still a challenge. It remains a future task for ALMA, with its the resolution and sensitivity advancements, to resolve these disks (including inner and outer disk differences), their gaseous surface density profiles, and the chemical signatures of changes in the inner dust structure. Only then will we be able to examine how  photodissociation, ionization, and freeze-out processes affect the surface density of the gas by comparison to resolved dust observations.

\begin{acknowledgements} 

We would like to thank Remo Tilanus and Tim van Kempen for help with the data collection and reduction. This research is supported through a VIDI grant from the Netherlands Organization for Scientific Research (NWO).

\end{acknowledgements}

\bibliographystyle{aa}
\bibliography{15411}

\begin{thebibliography}{90}
\expandafter\ifx\csname natexlab\endcsname\relax\def\natexlab#1{#1}\fi

\bibitem[{{Aikawa} {et~al.}(2002){Aikawa}, {van Zadelhoff}, {van Dishoeck}, \&
  {Herbst}}]{aikawa2002}
{Aikawa}, Y., {van Zadelhoff}, G.~J., {van Dishoeck}, E.~F., \& {Herbst}, E.
  2002, \aap, 386, 622

\bibitem[{{Akeson} {et~al.}(2005){Akeson}, {Boden}, {Monnier}, {Millan-Gabet},
  {Beichman}, {Beletic}, {Calvet}, {Hartmann}, {Hillenbrand}, {Koresko},
  {Sargent}, \& {Tannirkulam}}]{akeson2005}
{Akeson}, R.~L., {Boden}, A.~F., {Monnier}, J.~D., {et~al.} 2005, \apj, 635,
  1173

\bibitem[{{Andrews}(2007)}]{andrews2007phd}
{Andrews}, S.~M. 2007, PhD thesis, University of Hawai'i at Manoa

\bibitem[{{Andrews} \& {Williams}(2005)}]{andrews2005}
{Andrews}, S.~M. \& {Williams}, J.~P. 2005, \apj, 631, 1134

\bibitem[{{Appenzeller} {et~al.}(2005){Appenzeller}, {Bertout}, \&
  {Stahl}}]{appenzeller2005}
{Appenzeller}, I., {Bertout}, C., \& {Stahl}, O. 2005, \aap, 434, 1005

\bibitem[{{Basri} {et~al.}(1997){Basri}, {Johns-Krull}, \&
  {Mathieu}}]{basri1997}
{Basri}, G., {Johns-Krull}, C.~M., \& {Mathieu}, R.~D. 1997, \aj, 114, 781

\bibitem[{{Basri} {et~al.}(1991){Basri}, {Martin}, \& {Bertout}}]{basri1991}
{Basri}, G., {Martin}, E.~L., \& {Bertout}, C. 1991, \aap, 252, 625

\bibitem[{{Beckwith}(1996)}]{beckwith1996}
{Beckwith}, S.~V.~W. 1996, \nat, 383, 139

\bibitem[{{Beckwith} \& {Sargent}(1991)}]{beckwith1991}
{Beckwith}, S.~V.~W. \& {Sargent}, A.~I. 1991, \apj, 381, 250

\bibitem[{{Beckwith} {et~al.}(1990){Beckwith}, {Sargent}, {Chini}, \&
  {Guesten}}]{beckwith1990}
{Beckwith}, S.~V.~W., {Sargent}, A.~I., {Chini}, R.~S., \& {Guesten}, R. 1990,
  \aj, 99, 924

\bibitem[{{Bergin} {et~al.}(2003){Bergin}, {Calvet}, {D'Alessio}, \&
  {Herczeg}}]{bergin2003}
{Bergin}, E., {Calvet}, N., {D'Alessio}, P., \& {Herczeg}, G.~J. 2003, \apjl,
  591, L159

\bibitem[{{Bouvier} {et~al.}(1995){Bouvier}, {Covino}, {Kovo}, {Martin},
  {Matthews}, {Terranegra}, \& {Beck}}]{bouvier1995}
{Bouvier}, J., {Covino}, E., {Kovo}, O., {et~al.} 1995, \aap, 299, 89

\bibitem[{{Calvet} {et~al.}(2002){Calvet}, {D'Alessio}, {Hartmann}, {Wilner},
  {Walsh}, \& {Sitko}}]{calvet2002}
{Calvet}, N., {D'Alessio}, P., {Hartmann}, L., {et~al.} 2002, \apj, 568, 1008

\bibitem[{{Chapillon} {et~al.}(2008){Chapillon}, {Guilloteau}, {Dutrey}, \&
  {Pi{\'e}tu}}]{chapillon2008}
{Chapillon}, E., {Guilloteau}, S., {Dutrey}, A., \& {Pi{\'e}tu}, V. 2008, \aap,
  488, 565

\bibitem[{{Chiang} \& {Goldreich}(1997)}]{chiang1997}
{Chiang}, E.~I. \& {Goldreich}, P. 1997, \apj, 490, 368

\bibitem[{{Cieza} {et~al.}(2005){Cieza}, {Kessler-Silacci}, {Jaffe}, {Harvey},
  \& {Evans}}]{cieza2005}
{Cieza}, L.~A., {Kessler-Silacci}, J.~E., {Jaffe}, D.~T., {Harvey}, P.~M., \&
  {Evans}, II, N.~J. 2005, \apj, 635, 422

\bibitem[{{Damiani} {et~al.}(1995){Damiani}, {Micela}, {Sciortino}, \&
  {Harnden}}]{damiani1995a}
{Damiani}, F., {Micela}, G., {Sciortino}, S., \& {Harnden}, Jr., F.~R. 1995,
  \apj, 446, 331

\bibitem[{{Dullemond} \& {Dominik}(2004)}]{dullemond2004}
{Dullemond}, C.~P. \& {Dominik}, C. 2004, \aap, 421, 1075

\bibitem[{{Dutrey} {et~al.}(1996){Dutrey}, {Guilloteau}, {Duvert}, {Prato},
  {Simon}, {Schuster}, \& {Menard}}]{dutrey1996}
{Dutrey}, A., {Guilloteau}, S., {Duvert}, G., {et~al.} 1996, \aap, 309, 493

\bibitem[{{Dutrey} {et~al.}(1997){Dutrey}, {Guilloteau}, \&
  {Guelin}}]{dutrey1997}
{Dutrey}, A., {Guilloteau}, S., \& {Guelin}, M. 1997, \aap, 317, L55

\bibitem[{{Dutrey} {et~al.}(2007){Dutrey}, {Guilloteau}, \& {Ho}}]{dutrey2007}
{Dutrey}, A., {Guilloteau}, S., \& {Ho}, P. 2007, in Protostars and Planets V,
  ed. {B.~Reipurth, D.~Jewitt, \& K.~Keil}, 495--506

\bibitem[{{Eisl{\"o}ffel} \& {Mundt}(1998)}]{eisloffel1998}
{Eisl{\"o}ffel}, J. \& {Mundt}, R. 1998, \aj, 115, 1554

\bibitem[{{Fuente} {et~al.}(1993){Fuente}, {Martin-Pintado}, {Cernicharo}, \&
  {Bachiller}}]{fuente1993}
{Fuente}, A., {Martin-Pintado}, J., {Cernicharo}, J., \& {Bachiller}, R. 1993,
  \aap, 276, 473

\bibitem[{{Furlan} {et~al.}(2006){Furlan}, {Hartmann}, {Calvet}, {D'Alessio},
  {Franco-Hern{\'a}ndez}, {Forrest}, {Watson}, {Uchida}, {Sargent}, {Green},
  {Keller}, \& {Herter}}]{furlan2006}
{Furlan}, E., {Hartmann}, L., {Calvet}, N., {et~al.} 2006, \apjs, 165, 568

\bibitem[{{Glassgold} {et~al.}(2004){Glassgold}, {Najita}, \&
  {Igea}}]{glassgold2004}
{Glassgold}, A.~E., {Najita}, J., \& {Igea}, J. 2004, \apj, 615, 972

\bibitem[{{Greaves}(2004)}]{greaves2004}
{Greaves}, J.~S. 2004, \mnras, 351, L99

\bibitem[{{Greaves}(2005)}]{greaves2005}
{Greaves}, J.~S. 2005, \mnras, 364, L47

\bibitem[{{Greaves} \& {Church}(1996)}]{greaves1996}
{Greaves}, J.~S. \& {Church}, S.~E. 1996, \mnras, 283, 1179

\bibitem[{{G{\"u}del}(2008)}]{gudel2008a}
{G{\"u}del}, M. 2008, Astronomische Nachrichten, 329, 218

\bibitem[{{G{\"u}del} {et~al.}(2007){G{\"u}del}, {Briggs}, {Arzner}, {Audard},
  {Bouvier}, {Feigelson}, {Franciosini}, {Glauser}, {Grosso}, {Micela},
  {Monin}, {Montmerle}, {Padgett}, {Palla}, {Pillitteri}, {Rebull}, {Scelsi},
  {Silva}, {Skinner}, {Stelzer}, \& {Telleschi}}]{gudel2007a}
{G{\"u}del}, M., {Briggs}, K.~R., {Arzner}, K., {et~al.} 2007, \aap, 468, 353

\bibitem[{{G{\"u}del} {et~al.}(2008){G{\"u}del}, {Skinner}, {Audard}, {Briggs},
  \& {Cabrit}}]{gudel2008b}
{G{\"u}del}, M., {Skinner}, S.~L., {Audard}, M., {Briggs}, K.~R., \& {Cabrit},
  S. 2008, \aap, 478, 797

\bibitem[{{Guilloteau} {et~al.}(1999){Guilloteau}, {Dutrey}, \&
  {Simon}}]{guilloteau1999}
{Guilloteau}, S., {Dutrey}, A., \& {Simon}, M. 1999, \aap, 348, 570

\bibitem[{{Hartigan} {et~al.}(1995){Hartigan}, {Edwards}, \&
  {Ghandour}}]{hartigan1995}
{Hartigan}, P., {Edwards}, S., \& {Ghandour}, L. 1995, \apj, 452, 736

\bibitem[{{Hartmann} {et~al.}(1998){Hartmann}, {Calvet}, {Gullbring}, \&
  {D'Alessio}}]{hartmann1998}
{Hartmann}, L., {Calvet}, N., {Gullbring}, E., \& {D'Alessio}, P. 1998, \apj,
  495, 385

\bibitem[{{Herbig} \& {Bell}(1988)}]{herbig1988}
{Herbig}, G.~H. \& {Bell}, K.~R. 1988, {Third Catalog of Emission-Line Stars of
  the Orion Population : 3 : 1988}, ed. {Herbig, G.~H.~\& Bell, K.~R.}

\bibitem[{{Herbig} \& {Goodrich}(1986)}]{herbig1986}
{Herbig}, G.~H. \& {Goodrich}, R.~W. 1986, \apj, 309, 294

\bibitem[{{Herczeg} \& {Hillenbrand}(2008)}]{herczeg2008}
{Herczeg}, G.~J. \& {Hillenbrand}, L.~A. 2008, \apj, 681, 594

\bibitem[{{Hessman} \& {Guenther}(1997)}]{hessman1997}
{Hessman}, F.~V. \& {Guenther}, E.~W. 1997, \aap, 321, 497

\bibitem[{{Hogerheijde} {et~al.}(1995){Hogerheijde}, {Jansen}, \& {van
  Dishoeck}}]{hogerheijde1995a}
{Hogerheijde}, M.~R., {Jansen}, D.~J., \& {van Dishoeck}, E.~F. 1995, \aap,
  294, 792

\bibitem[{{Hogerheijde} \& {Sandell}(2000)}]{hogerheijde2000a}
{Hogerheijde}, M.~R. \& {Sandell}, G. 2000, \apj, 534, 880

\bibitem[{{Hogerheijde} \& {van der Tak}(2000)}]{hogerheijde2000b}
{Hogerheijde}, M.~R. \& {van der Tak}, F.~F.~S. 2000, \aap, 362, 697

\bibitem[{{Hogerheijde} {et~al.}(1998){Hogerheijde}, {van Dishoeck}, {Blake},
  \& {van Langevelde}}]{hogerheijde1998}
{Hogerheijde}, M.~R., {van Dishoeck}, E.~F., {Blake}, G.~A., \& {van
  Langevelde}, H.~J. 1998, \apj, 502, 315

\bibitem[{{Honda} {et~al.}(2006){Honda}, {Kataza}, {Okamoto}, {Yamashita},
  {Min}, {Miyata}, {Sako}, {Fujiyoshi}, {Sakon}, \& {Onaka}}]{honda2006}
{Honda}, M., {Kataza}, H., {Okamoto}, Y.~K., {et~al.} 2006, \apj, 646, 1024

\bibitem[{{Hughes} {et~al.}(2008){Hughes}, {Wilner}, {Qi}, \&
  {Hogerheijde}}]{hughes2008}
{Hughes}, A.~M., {Wilner}, D.~J., {Qi}, C., \& {Hogerheijde}, M.~R. 2008, \apj,
  678, 1119

\bibitem[{{Isella} {et~al.}(2009){Isella}, {Carpenter}, \&
  {Sargent}}]{isella2009}
{Isella}, A., {Carpenter}, J.~M., \& {Sargent}, A.~I. 2009, \apj, 701, 260

\bibitem[{{J{\o}rgensen} {et~al.}(2004){J{\o}rgensen}, {Sch{\"o}ier}, \& {van
  Dishoeck}}]{jorgensen2004}
{J{\o}rgensen}, J.~K., {Sch{\"o}ier}, F.~L., \& {van Dishoeck}, E.~F. 2004,
  \aap, 416, 603

\bibitem[{{Kastner} {et~al.}(2008{\natexlab{a}}){Kastner}, {Zuckerman}, \&
  {Forveille}}]{kastner2008a}
{Kastner}, J.~H., {Zuckerman}, B., \& {Forveille}, T. 2008{\natexlab{a}}, \aap,
  486, 239

\bibitem[{{Kastner} {et~al.}(2008{\natexlab{b}}){Kastner}, {Zuckerman},
  {Hily-Blant}, \& {Forveille}}]{kastner2008b}
{Kastner}, J.~H., {Zuckerman}, B., {Hily-Blant}, P., \& {Forveille}, T.
  2008{\natexlab{b}}, \aap, 492, 469

\bibitem[{{Kastner} {et~al.}(1997){Kastner}, {Zuckerman}, {Weintraub}, \&
  {Forveille}}]{kastner1997}
{Kastner}, J.~H., {Zuckerman}, B., {Weintraub}, D.~A., \& {Forveille}, T. 1997,
  Science, 277, 67

\bibitem[{{Kenyon} {et~al.}(1994){Kenyon}, {Dobrzycka}, \&
  {Hartmann}}]{kenyon1994}
{Kenyon}, S.~J., {Dobrzycka}, D., \& {Hartmann}, L. 1994, \aj, 108, 1872

\bibitem[{{Kenyon} \& {Hartmann}(1995)}]{kenyon1995}
{Kenyon}, S.~J. \& {Hartmann}, L. 1995, \apjs, 101, 117

\bibitem[{{Kessler-Silacci}(2004)}]{kesslerphd2004}
{Kessler-Silacci}, J. 2004, PhD thesis, AA(CALIFORNIA INSTITUTE OF TECHNOLOGY)

\bibitem[{{Kitamura} {et~al.}(1996{\natexlab{a}}){Kitamura}, {Kawabe}, \&
  {Saito}}]{kitamura1996a}
{Kitamura}, Y., {Kawabe}, R., \& {Saito}, M. 1996{\natexlab{a}}, \apj, 457, 277

\bibitem[{{Kitamura} {et~al.}(1996{\natexlab{b}}){Kitamura}, {Kawabe}, \&
  {Saito}}]{kitamura1996b}
{Kitamura}, Y., {Kawabe}, R., \& {Saito}, M. 1996{\natexlab{b}}, \apjl, 465,
  L137+

\bibitem[{{Koerner} \& {Sargent}(1995)}]{koerner1995}
{Koerner}, D.~W. \& {Sargent}, A.~I. 1995, \aj, 109, 2138

\bibitem[{{Lepp} \& {Dalgarno}(1996)}]{lepp1996}
{Lepp}, S. \& {Dalgarno}, A. 1996, \aap, 306, L21

\bibitem[{{Luhman} {et~al.}(2010){Luhman}, {Allen}, {Espaillat}, {Hartmann}, \&
  {Calvet}}]{luhman2010}
{Luhman}, K.~L., {Allen}, P.~R., {Espaillat}, C., {Hartmann}, L., \& {Calvet},
  N. 2010, \apjs, 186, 111

\bibitem[{{Mannings} \& {Sargent}(1997)}]{mannings1997}
{Mannings}, V. \& {Sargent}, A.~I. 1997, \apj, 490, 792

\bibitem[{{Mannings} \& {Sargent}(2000)}]{mannings2000}
{Mannings}, V. \& {Sargent}, A.~I. 2000, \apj, 529, 391

\bibitem[{{Mer{\'{\i}}n} {et~al.}(2004){Mer{\'{\i}}n}, {Montesinos}, {Eiroa},
  {Solano}, {Mora}, {D'Alessio}, {Calvet}, {Oudmaijer}, {de Winter}, {Davies},
  {Harris}, {Cameron}, {Deeg}, {Ferlet}, {Garz{\'o}n}, {Grady}, {Horne},
  {Miranda}, {Palacios}, {Penny}, {Quirrenbach}, {Rauer}, {Schneider}, \&
  {Wesselius}}]{merin2004}
{Mer{\'{\i}}n}, B., {Montesinos}, B., {Eiroa}, C., {et~al.} 2004, \aap, 419,
  301

\bibitem[{{Mitchell} {et~al.}(1994){Mitchell}, {Hasegawa}, {Dent}, \&
  {Matthews}}]{mitchell1994}
{Mitchell}, G.~F., {Hasegawa}, T.~I., {Dent}, W.~R.~F., \& {Matthews}, H.~E.
  1994, \apjl, 436, L177

\bibitem[{{Mohanty} {et~al.}(2005){Mohanty}, {Jayawardhana}, \&
  {Basri}}]{mohanty2005}
{Mohanty}, S., {Jayawardhana}, R., \& {Basri}, G. 2005, \apj, 626, 498

\bibitem[{{Muzerolle} {et~al.}(2003){Muzerolle}, {Calvet}, {Hartmann}, \&
  {D'Alessio}}]{muzerolle2003}
{Muzerolle}, J., {Calvet}, N., {Hartmann}, L., \& {D'Alessio}, P. 2003, \apjl,
  597, L149

\bibitem[{{Natta} {et~al.}(2007){Natta}, {Testi}, {Calvet}, {Henning},
  {Waters}, \& {Wilner}}]{natta2007}
{Natta}, A., {Testi}, L., {Calvet}, N., {et~al.} 2007, Protostars and Planets
  V, 767

\bibitem[{{Palla} \& {Stahler}(2002)}]{palla2002}
{Palla}, F. \& {Stahler}, S.~W. 2002, \apj, 581, 1194

\bibitem[{{Pani{\'c}} {et~al.}(2008){Pani{\'c}}, {Hogerheijde}, {Wilner}, \&
  {Qi}}]{panic2008}
{Pani{\'c}}, O., {Hogerheijde}, M.~R., {Wilner}, D., \& {Qi}, C. 2008, \aap,
  491, 219

\bibitem[{{Pi{\'e}tu} {et~al.}(2007){Pi{\'e}tu}, {Dutrey}, \&
  {Guilloteau}}]{pietu2007}
{Pi{\'e}tu}, V., {Dutrey}, A., \& {Guilloteau}, S. 2007, \aap, 467, 163

\bibitem[{{Rebull} {et~al.}(2010){Rebull}, {Padgett}, {McCabe}, {Hillenbrand},
  {Stapelfeldt}, {Noriega-Crespo}, {Carey}, {Brooke}, {Huard}, {Terebey},
  {Audard}, {Monin}, {Fukagawa}, {G{\"u}del}, {Knapp}, {Menard}, {Allen},
  {Angione}, {Baldovin-Saavedra}, {Bouvier}, {Briggs}, {Dougados}, {Evans},
  {Flagey}, {Guieu}, {Grosso}, {Glauser}, {Harvey}, {Hines}, {Latter},
  {Skinner}, {Strom}, {Tromp}, \& {Wolf}}]{rebull2010}
{Rebull}, L.~M., {Padgett}, D.~L., {McCabe}, C., {et~al.} 2010, \apjs, 186, 259

\bibitem[{{Robitaille} {et~al.}(2007){Robitaille}, {Whitney}, {Indebetouw}, \&
  {Wood}}]{robitaille2007}
{Robitaille}, T.~P., {Whitney}, B.~A., {Indebetouw}, R., \& {Wood}, K. 2007,
  \apjs, 169, 328

\bibitem[{{Robitaille} {et~al.}(2006){Robitaille}, {Whitney}, {Indebetouw},
  {Wood}, \& {Denzmore}}]{robitaille2006}
{Robitaille}, T.~P., {Whitney}, B.~A., {Indebetouw}, R., {Wood}, K., \&
  {Denzmore}, P. 2006, \apjs, 167, 256

\bibitem[{{Rodmann} {et~al.}(2006){Rodmann}, {Henning}, {Chandler}, {Mundy}, \&
  {Wilner}}]{rodmann2006}
{Rodmann}, J., {Henning}, T., {Chandler}, C.~J., {Mundy}, L.~G., \& {Wilner},
  D.~J. 2006, \aap, 446, 211

\bibitem[{{Schaefer} {et~al.}(2009){Schaefer}, {Dutrey}, {Guilloteau}, {Simon},
  \& {White}}]{schaefer2009}
{Schaefer}, G.~H., {Dutrey}, A., {Guilloteau}, S., {Simon}, M., \& {White},
  R.~J. 2009, \apj, 701, 698

\bibitem[{{Schreyer} {et~al.}(2008){Schreyer}, {Guilloteau}, {Semenov},
  {Bacmann}, {Chapillon}, {Dutrey}, {Gueth}, {Henning}, {Hersant}, {Launhardt},
  {Pety}, \& {Pi{\'e}tu}}]{schreyer2008}
{Schreyer}, K., {Guilloteau}, S., {Semenov}, D., {et~al.} 2008, \aap, 491, 821

\bibitem[{{Schuster} {et~al.}(1993){Schuster}, {Harris}, {Anderson}, \&
  {Russell}}]{schuster1993}
{Schuster}, K.~F., {Harris}, A.~I., {Anderson}, N., \& {Russell}, A.~P.~G.
  1993, \apjl, 412, L67

\bibitem[{{Simon} {et~al.}(2000){Simon}, {Dutrey}, \& {Guilloteau}}]{simon2000}
{Simon}, M., {Dutrey}, A., \& {Guilloteau}, S. 2000, \apj, 545, 1034

\bibitem[{{Tamura} {et~al.}(1999){Tamura}, {Hough}, {Greaves}, {Morino},
  {Chrysostomou}, {Holland}, \& {Momose}}]{tamura1999}
{Tamura}, M., {Hough}, J.~H., {Greaves}, J.~S., {et~al.} 1999, \apj, 525, 832

\bibitem[{{Terzieva} \& {Herbst}(1998)}]{terzieva1998}
{Terzieva}, R. \& {Herbst}, E. 1998, \apj, 501, 207

\bibitem[{{Testi} {et~al.}(2002){Testi}, {Bacciotti}, {Sargent}, {Ray}, \&
  {Eisl{\"o}ffel}}]{testi2002}
{Testi}, L., {Bacciotti}, F., {Sargent}, A.~I., {Ray}, T.~P., \&
  {Eisl{\"o}ffel}, J. 2002, \aap, 394, L31

\bibitem[{{Testi} {et~al.}(2003){Testi}, {Natta}, {Shepherd}, \&
  {Wilner}}]{testi2003}
{Testi}, L., {Natta}, A., {Shepherd}, D.~S., \& {Wilner}, D.~J. 2003, \aap,
  403, 323

\bibitem[{{Thi} {et~al.}(2004){Thi}, {van Zadelhoff}, \& {van
  Dishoeck}}]{thi2004}
{Thi}, W., {van Zadelhoff}, G., \& {van Dishoeck}, E.~F. 2004, \aap, 425, 955

\bibitem[{{Thi} {et~al.}(2001){Thi}, {van Dishoeck}, {Blake}, {van Zadelhoff},
  {Horn}, {Becklin}, {Mannings}, {Sargent}, {van den Ancker}, {Natta}, \&
  {Kessler}}]{thi2001}
{Thi}, W.~F., {van Dishoeck}, E.~F., {Blake}, G.~A., {et~al.} 2001, \apj, 561,
  1074

\bibitem[{{van der Tak} {et~al.}(2007){van der Tak}, {Black}, {Sch{\"o}ier},
  {Jansen}, \& {van Dishoeck}}]{vandertak2007}
{van der Tak}, F.~F.~S., {Black}, J.~H., {Sch{\"o}ier}, F.~L., {Jansen}, D.~J.,
  \& {van Dishoeck}, E.~F. 2007, \aap, 468, 627

\bibitem[{{van Kempen} {et~al.}(2007){van Kempen}, {van Dishoeck}, {Brinch}, \&
  {Hogerheijde}}]{kempen2007}
{van Kempen}, T.~A., {van Dishoeck}, E.~F., {Brinch}, C., \& {Hogerheijde},
  M.~R. 2007, \aap, 461, 983

\bibitem[{{van Zadelhoff} {et~al.}(2003){van Zadelhoff}, {Aikawa},
  {Hogerheijde}, \& {van Dishoeck}}]{zadelhoff2003}
{van Zadelhoff}, G.-J., {Aikawa}, Y., {Hogerheijde}, M.~R., \& {van Dishoeck},
  E.~F. 2003, \aap, 397, 789

\bibitem[{{Vinkovi{\'c}} \& {Jurki{\'c}}(2007)}]{vinkovic2007}
{Vinkovi{\'c}}, D. \& {Jurki{\'c}}, T. 2007, \apj, 658, 462

\bibitem[{{Weidenschilling}(1997)}]{weidenschilling1997}
{Weidenschilling}, S.~J. 1997, Icarus, 127, 290

\bibitem[{{White} \& {Ghez}(2001)}]{white2001}
{White}, R.~J. \& {Ghez}, A.~M. 2001, \apj, 556, 265

\bibitem[{{White} \& {Hillenbrand}(2004)}]{white2004}
{White}, R.~J. \& {Hillenbrand}, L.~A. 2004, \apj, 616, 998

\bibitem[{{Whitney} {et~al.}(2003){Whitney}, {Wood}, {Bjorkman}, \&
  {Wolff}}]{whitney2003a}
{Whitney}, B.~A., {Wood}, K., {Bjorkman}, J.~E., \& {Wolff}, M.~J. 2003, \apj,
  591, 1049

\bibitem[{{Yun} {et~al.}(1999){Yun}, {Moreira}, {Afonso}, \&
  {Clemens}}]{yun1999}
{Yun}, J.~L., {Moreira}, M.~C., {Afonso}, J.~M., \& {Clemens}, D.~P. 1999, \aj,
  118, 990

\end{thebibliography}


\Online

\end{document}